\documentstyle[12pt,epsfig,psfig]{article}
        \newdimen\eqskip
        \newdimen\txtskip
        \baselineskip=\txtskip
\begin{document}

  \newcommand{\ccaption}[2]{
    \begin{center}
    \parbox{0.85\textwidth}{
      \caption[#1]{\small{{#2}}}
      }
    \end{center}
    }
\newcommand{\BS}{\bigskip}
\def    \be             {\begin{equation}}
\def    \ee             {\end{equation}}
\def    \ba             {\begin{eqnarray}}
\def    \ea             {\end{eqnarray}}
\def    \nn             {\nonumber}
\def    \=              {\;=\;}
\def    \frac           #1#2{{#1 \over #2}}
\def    \ret            {\\[\eqskip]}
\def    \ie             {{\em i.e.\/} }
\def    \eg             {{\em e.g.\/} }
\def    \lsim           {\raisebox{-3pt}{$\>\stackrel{<}{\scriptstyle\sim}\>$}}
\def    \bentarrow      {\:\raisebox{1.1ex}{\rlap{$\vert$}}\!\rightarrow}
\def    \rd             {{\mathrm d}}    
\def    \Im             {{\mathrm{Im}}}  
\newcommand{\bra}[1]{{\langle #1 |}}
\newcommand{\ket}[1]{{| #1 \rangle}}

\def    \kev            {\mbox{$\mathrm{keV}$}}
\def    \mev            {\mbox{$\mathrm{MeV}$}}
\def    \gev            {\mbox{$\mathrm{GeV}$}}


\def    \mq             {\mbox{$m_Q$}}  
\def    \mqq            {\mbox{$m_{Q\bar Q}$}}
\def    \mqqsq          {\mbox{$m^2_{Q\bar Q}$}}
\def    \pt             {\mbox{$p_T$}}
\def    \ptsq           {\mbox{$p^2_T$}}

\def    \o              {\ifmmode {\cal{O}} \else ${\cal{Q}}$ \fi}
\def    \q              {\ifmmode {\cal{Q}} \else ${\cal{Q}}$ \fi}
\def    \oo              {\ifmmode {\cal{O}} \else 
                            $\overline{\cal{O}}$ \fi}
\def    \ups            {\ifmmode \Upsilon \else $\Upsilon$ \fi}
\def    \oneSzero       {\ifmmode {^1S_0} \else $^1S_0$ \fi}
\def    \threeSone      {\ifmmode {^3S_1} \else $^3S_1$ \fi}
\def    \onePone        {\ifmmode {^1P_1} \else $^1P_1$ \fi}
\def    \threePJ        {\ifmmode {^3P_J} \else $^3P_J$ \fi}
\def    \threePzero     {\ifmmode {^3P_0} \else $^3P_0$ \fi}
\def    \threePone      {\ifmmode {^3P_1} \else $^3P_1$ \fi}
\def    \threePtwo      {\ifmmode {^3P_2} \else $^3P_2$ \fi}

\def    \heightp         {\mbox{$H_8^{\prime}$}}
\def    \vevpsi         {\mbox{$\langle {\cal O}_8^{\psi}(^3S_1) \rangle$}}
\def    \vevpsp         {\mbox{$\langle {\cal O}_8^{\psp}(^3S_1) \rangle$}}
\newcommand     \MSB            {\ifmmode {\overline{\rm MS}} \else 
                                 $\overline{\rm MS}$  \fi}
\def    \muf            {\mbox{$\mu_{\rm F}$}}
\def    \mufsq          {\mbox{$\mu^2_{\rm F}$}}
\def    \mur            {{\mbox{$\mu_{\rm R}$}}}
\def    \mursq          {\mbox{$\mu^2_{\rm R}$}}
\def    \mul            {{\mu_\Lambda}}
\def    \mulsq          {\mbox{$\mu^2_\Lambda$}}

\def    \as             {\mbox{$\alpha_s$}}
\def    \asb            {\mbox{$\alpha_s^{(b)}$}}
\def    \assq           {\mbox{$\alpha_s^2$}}
\def    \ascube         {\mbox{$\alpha_s^3$}}
\def    \asfour         {\mbox{$\alpha_s^4$}}
\def    \asfive         {\mbox{$\alpha_s^5$}}
\def    \alp            {\mbox{$e^2_Q \alpha$}}

\def    \eps            {\ifmmode \epsilon \else $\epsilon$ \fi}
\def    \epsbar         {\ifmmode \bar\epsilon \else $\bar\epsilon$ \fi}
\def    \epsir          {\ifmmode \epsilon_{\rm IR} \else $\epsilon_{\rm IR}$ \fi}
\def    \epsuv          {\ifmmode \epsilon_{\rm UV} \else $\epsilon_{\rm UV}$ \fi}


\def    \Atot           {{\rm A_{tot}}}
\def    \aeight         {A^{[8]}}
\def    \aeightb        {A^{[8]}_{\rm Born}}
\def    \aeights        {A^{[8]}_{\rm soft}}
\def    \ovaeight         {\overline{A^{[8]}}}
\def    \ovaeightb        {\overline{A^{[8]}}_{\rm Born}}
\def    \ovaeights        {\overline{A^{[8]}}_{\rm soft}}
\def    \aone           {A^{[1]}}
\def    \asins          {A^{[1]}_{\rm soft}}
\def    \aoneb          {A^{[1]}_{\rm Born}[^3P_J]}
\def    \aones          {A^{[1]}_{\rm soft}[^3P_J]}
\def    \aoneo          {A^{[8]}_{\rm soft}[^3P_J]}
\def    \ovaone           {\overline{A^{[1]}}}
\def    \ovaoneb          {\overline{A^{[1]}}_{\rm Born}[^3P_J]}
\def    \ovaones          {\overline{A^{[1]}}_{\rm soft}[^3P_J]}
\def    \ovaoneo          {\overline{A^{[8]}}_{\rm soft}[^3P_J]}

\def    \C              {\frac{\Phi_{(2)}}{2 M} \, \frac{N}{K}}
\def    \Cggg           {\frac{1}{2M} \frac{\Phi_{(2)}}{3!}\frac{N}{K}}
\def    \Cgggamma       {\frac{1}{2M} \frac{\Phi_{(2)}}{2!}\frac{N}{K}}
\def    \Cgggad         {\frac{ \Phi_{(2)} }{2M}\frac{N}{K}}
\def    \m              {\mbox{${\cal{M}}$}}
\def    \mbar           {\mbox{${\overline{\cal{M}}}$}}
\def    \mborn          {\mbox{${\cal{M}}_{\rm Born}$}}
\def    \gborn          {\mbox{$\Gamma_{\rm Born}$}}
\def    \gbh            {\mbox{$\Gamma_{\rm Born}$}}
\def    \gb            {\mbox{$\Gamma_{\rm Born}$}}
\def    \sborn          {\mbox{$\sigma_{\rm Born}$}}
\def    \sborno         {\mbox{$\sigma^0_{\rm Born}$}}
\def    \sborno         {\mbox{$\sigma_0$}}
\def    \pgg#1          {P_{gg}(#1)}
\def    \cpgq#1         {{\cal{P}}_{gq}(#1)}
\def    \cpgg#1         {{\cal{P}}_{gg}(#1)}
\def    \cpqq#1         {{\cal{P}}_{qq}(#1)}
\def    \cpqgamma#1     {{\cal{P}}_{q\gamma}(#1)}
\def    \dk             {\mbox{${\cal D}_k$}}
\def    \fk             {\mbox{$f_k$}}
\def    \sp#1#2         {#1#2}       
\def    \eik#1          { \frac{#1 \epsilon_c}{#1 k} }
\def    \eikg#1          { \frac{#1 \epsilon}{#1 k} }
\def    \hone   {\mbox{$H_1$}}
\def    \height {\mbox{$H_8$}}

\def\jpsi{\mbox{$J\!/\!\psi$}}
\def\chic{\mbox{$\chi_c$}}
\def\chij{\mbox{$\chi_J$}}
\def\chicj{\mbox{$\chi_{cJ}$}}
\def\psp {\mbox{$\psi'$}}
\def\mups {\mbox{$M_\Upsilon$}}

\def \oacube {\mbox{$ O(\alpha_s^3)$}}
\def \oatwo {\mbox{$ O(\alpha_s^2)$}}
\def \oas   {\mbox{$ O(\alpha_s)$}}

\def \chiz {\mbox{$\chi_{0}$}}
\def \chio {\mbox{$\chi_{1}$}}
\def \chit {\mbox{$\chi_{2}$}}

\def \chiqz {\mbox{$\chi_{Q0}$}}
\def \chiqo {\mbox{$\chi_{Q1}$}}
\def \chiqt {\mbox{$\chi_{Q2}$}}

\def \chicz {\mbox{$\chi_{c0}$}}
\def \chico {\mbox{$\chi_{c1}$}}
\def \chict {\mbox{$\chi_{c2}$}}

\def \chibz {\mbox{$\chi_{b0}$}}
\def \chibo {\mbox{$\chi_{b1}$}}
\def \chibt {\mbox{$\chi_{b2}$}}

\def \QQ {Q \overline Q}
\def \qq {\mbox{$q \overline q$}}
\def \cc {\mbox{$c \overline c$}}

\def\lqcd{\mbox{$\Lambda_{QCD}$}}

\def \chizgg {\mbox{$\Gamma(\chiz \to \gamma\gamma)$}}
\def \chitgg {\mbox{$\Gamma(\chit \to \gamma\gamma)$}}
\def \chijgg {\mbox{$\Gamma(\chi_J \to \gamma\gamma)$}}
\def \chizlh {\mbox{$\Gamma(\chiz \to LH)$}}
\def \chiolh {\mbox{$\Gamma(\chio \to LH)$}}
\def \chitolh {\mbox{$\Gamma(\chij \to LH)$}}
\def \chijlh {\mbox{$\Gamma(\chi_J \to LH)$}}
\def \chijqqg{\mbox{$\Gamma(\chi_J \to q \overline q g)$}}
\def \lh{{\mathrm LH}}
\def \lp{{\mathrm LP}}

\def \ebind {\mbox{$E_{\mathrm bind}$}}
\def \rprime {\mbox{${\cal R}^{\prime}_{1P}(0)$}}
\def \rprimes {\mbox{$\vert {\cal R}^{\prime}_{1P}(0) \vert^2$}}
\def \rr {\mbox{${\cal R}_S(0)$}}
\def \rros {\mbox{$\vert{\cal R}_{1S}(0)\vert^2$}}
\def \rrts {\mbox{$\vert{\cal R}_{2S}(0)\vert^2$}}
\def \rrns {\mbox{$\vert{\cal R}_{nS}(0)\vert^2$}}
\def \ei {\mbox{$ \epsilon_{ \!\!\mbox{ \tiny{IR} } } $}}
\def \eu {\mbox{$ \epsilon_{ \!\!\mbox{ \tiny{UV} } } $}}
\def \s0 {\mbox{$\sigma_{0} $}}
\def \se {\mbox{$\sigma_{0}(\epsilon) $}}
\def \ep {\mbox{$\epsilon $}}

\def \cf {\mbox{$ C_F $}}
\def \ca {\mbox{$ C_A $}}
\def \caf {\mbox{$ C_F-\frac{1}{2}C_A $}}
\def \tf {\mbox{$ T_F $}}
\def \nf {\mbox{$n_f$}}
\def \da {\mbox{$ D_A $}}
\def \Bf {\mbox{$ B_F $}}
\def \df {\mbox{$ D_F $}}

\def \fe{\mbox{$f(\ep)$}}
\def \de{\mbox{$F(\ep)$}}
\def \feps#1 {f_{\epsilon}(#1)}

\def\der{\mbox{$\stackrel{\leftrightarrow}{\bf D}$}}
\def\nder{\mbox{$\stackrel{\leftrightarrow}{D}$}}
\def\tr{{\mathrm Tr}}

\def\opchizh {\mbox{$\langle 0\vert {\cal O}_8^{\psi}(^3P_0)\vert 0 \rangle$}}
\def\opchioh {\mbox{$\langle 0\vert {\cal O}_8^{\psi}(^3P_1)\vert 0 \rangle$}}
\def\opchith {\mbox{$\langle 0\vert {\cal O}_8^{\psi}(^3P_2)\vert 0 \rangle$}}

\def\opchijh {\mbox{$\langle 0\vert {\cal O}_8^{\psi}(^3P_J)\vert 0 \rangle$}}
\def\opetah  {\mbox{$\langle 0\vert {\cal O}_8^{\psi}(^1S_0)\vert 0 \rangle$}}
\def\oppsih  {\mbox{$\langle 0\vert {\cal O}_8^{\psi}(^3S_1)\vert 0 \rangle$}}
\def\oppsis  {\mbox{$\langle 0\vert {\cal O}_1^{\psi}(^3S_1)\vert 0 \rangle$}}
\def\pppsisq  {\mbox{$\langle 0\vert {\cal P}_1^{\psi_Q}(^3S_1)\vert 0 \rangle$}}

\def\opchijhq {\mbox{$\langle 0\vert {\cal O}_8^{\psi_Q}(^3P_J)\vert 0 \rangle$}}
\def\opetahq  {\mbox{$\langle 0\vert {\cal O}_8^{\psi_Q}(^1S_0)\vert 0 \rangle$}}
\def\oppsihq  {\mbox{$\langle 0\vert {\cal O}_8^{\psi_Q}(^3S_1)\vert 0 \rangle$}}
\def\oppsisq  {\mbox{$\langle 0\vert {\cal O}_1^{\psi_Q}(^3S_1)\vert 0 \rangle$}}
\def\ppchijh  {\mbox{$\langle 0\vert {\cal O}_8^{\tiny\chij}(^3S_1)\vert 0 \rangle$}}
\def\opchizhq {\mbox{$\langle 0\vert {\cal O}_8^{\psi_Q}(^3P_0)\vert 0 \rangle$}}
\def\opchiohq {\mbox{$\langle 0\vert {\cal O}_8^{\psi_Q}(^3P_1)\vert 0 \rangle$}}
\def\opchithq {\mbox{$\langle 0\vert {\cal O}_8^{\psi_Q}(^3P_2)\vert 0 \rangle$}}

\def\ppchizh  {\mbox{$\langle 0\vert {\cal O}_8^{\tiny\chiz}(^3S_1)\vert 0 \rangle$}}
\def\ppchioh  {\mbox{$\langle 0\vert {\cal O}_8^{\tiny\chio}(^3S_1)\vert 0 \rangle$}}
\def\ppchith  {\mbox{$\langle 0\vert {\cal O}_8^{\tiny\chit}(^3S_1)\vert 0 \rangle$}}

\def\ppchiczh  {\mbox{$\langle 0\vert {\cal O}_8^{\tiny\chicz}(^3S_1)\vert 0 \rangle$}}
\def\ppchicoh  {\mbox{$\langle 0\vert {\cal O}_8^{\tiny\chico}(^3S_1)\vert 0 \rangle$}}
\def\ppchicth  {\mbox{$\langle 0\vert {\cal O}_8^{\tiny\chict}(^3S_1)\vert 0 \rangle$}}

\def\oppsp  {\mbox{$\langle 0\vert {\cal O}_8^{\psi^{\prime}}(^3S_1)\vert 0 \rangle$}}
\def\oppsps  {\mbox{$\langle 0\vert {\cal O}_1^{\psi^{\prime}}(^3S_1)\vert 0 \rangle$}}

\def\spectrc {\mbox{$^{2S+1}L^{[c]}_J$}}
\def\spectr {\mbox{$^{2S+1}L_J$}}
\def\spectrs {\mbox{$^{2S+1}L_J^{[1]}$}}
\def\spectrh {\mbox{$^{2S+1}L_J^{[8]}$}}
\def\szh    {\mbox{$\sigma_0^{H}$}}
\def\sbh    {\mbox{$\sigma_{\rm Born}^{H}$}}
\def\szpsi {\mbox{$\sigma_0^{\psi}$}}
\def\szpsiq {\mbox{$\sigma_0^{\psi_Q}$}}
\def\spectro {{^{2S+1}L^{[8]}_J}}

\def\etah {\mbox{$^1S_0^{[8]}$}}
\def\etas {\mbox{$^1S_0^{[1]}$}}
\def\psih {\mbox{$^3S_1^{[8]}$}}
\def\psis {\mbox{$^3S_1^{[1]}$}}
\def\chizh {\mbox{$^3P_0^{[8]}$}}
\def\chioh {\mbox{$^3P_1^{[8]}$}}
\def\chith {\mbox{$^3P_2^{[8]}$}}
\def\chijh {\mbox{$^3P_J^{[8]}$}}
\def\chizs {\mbox{$^3P_0^{[1]}$}}
\def\chios {\mbox{$^3P_1^{[1]}$}}
\def\chits {\mbox{$^3P_2^{[1]}$}}
\def\chijs {\mbox{$^3P_J^{[1]}$}}
\def\chijo {\mbox{$^3P_J^{[8]}$}}
\def\opchizs {\mbox{$\langle 0\vert {\cal O}_1^{\tiny\chiz}(^3P_0)\vert 0 \rangle$}}
\def\opchios {\mbox{$\langle 0\vert {\cal O}_1^{\tiny\chio}(^3P_1)\vert 0 \rangle$}}
\def\opchits {\mbox{$\langle 0\vert {\cal O}_1^{\tiny\chit}(^3P_2)\vert 0 \rangle$}}
\def\opchijs {\mbox{$\langle 0\vert {\cal O}_1^{\tiny\chij}(^3P_J)\vert 0 \rangle$}}
\def\opchizh {\mbox{$\langle 0\vert {\cal O}_8^{\psi}(^3P_0)\vert 0 \rangle$}}
\def\opchioh {\mbox{$\langle 0\vert {\cal O}_8^{\psi}(^3P_1)\vert 0 \rangle$}}
\def\opchith {\mbox{$\langle 0\vert {\cal O}_8^{\psi}(^3P_2)\vert 0 \rangle$}}
\def\opchijh {\mbox{$\langle 0\vert {\cal O}_8^{\psi}(^3P_J)\vert 0 \rangle$}}
\def\opetah  {\mbox{$\langle 0\vert {\cal O}_8^{\psi}(^1S_0)\vert 0 \rangle$}}
\def\oppsih  {\mbox{$\langle 0\vert {\cal O}_8^{\psi}(^3S_1)\vert 0 \rangle$}}
\def\oppsis  {\mbox{$\langle 0\vert {\cal O}_1^{\psi}(^3S_1)\vert 0 \rangle$}}
\def\ophtpzs {\mbox{$\langle 0\vert {\cal O}_1^{H}(^3P_0)\vert 0 \rangle$}}
\def\ophtpos {\mbox{$\langle 0\vert {\cal O}_1^{H}(^3P_1)\vert 0 \rangle$}}
\def\ophtpts {\mbox{$\langle 0\vert {\cal O}_1^{H}(^3P_2)\vert 0 \rangle$}}
\def\ophtpjs {\mbox{$\langle 0\vert {\cal O}_1^{H}(^3P_J)\vert 0 \rangle$}}
\def\ophoszs {\mbox{$\langle 0\vert {\cal O}_1^{H}(^1S_0)\vert 0 \rangle$}}
\def\ophtsos {\mbox{$\langle 0\vert {\cal O}_1^{H}(^3S_1)\vert 0 \rangle$}}
\def\ophtpzo {\mbox{$\langle 0\vert {\cal O}_8^{H}(^3P_0)\vert 0 \rangle$}}
\def\ophtpoo {\mbox{$\langle 0\vert {\cal O}_8^{H}(^3P_1)\vert 0 \rangle$}}
\def\ophtpto {\mbox{$\langle 0\vert {\cal O}_8^{H}(^3P_2)\vert 0 \rangle$}}
\def\ophtpjo {\mbox{$\langle 0\vert {\cal O}_8^{H}(^3P_J)\vert 0 \rangle$}}
\def\ophoszo {\mbox{$\langle 0\vert {\cal O}_8^{H}(^1S_0)\vert 0 \rangle$}}
\def\ophtsoo {\mbox{$\langle 0\vert {\cal O}_8^{H}(^3S_1)\vert 0 \rangle$}}

\def\opchih {\mbox{$\langle 0\vert{\cal O}_8^{\tiny\chij}(^3S_1)\vert 0\rangle$}}

\def\b0{\mbox{$b_0$}}
\def\dsh  {\mbox{$\sigma^{H} $}}  
\def\dspq  {\mbox{$d\sigma^{\psi_Q} $}}
\def\dsp  {\mbox{$d\sigma^{\psi_Q} $}}
\def\qf     {\mbox{$\mu^2_{\rm F}$}}
\def\asopi{\mbox{$\frac{\as}{\pi}$}}
\def\szchiz{\mbox{$\sigma_0^{\tiny\chiz}$}}
\def\szchio{\mbox{$\sigma_0^{\tiny\chio}$}}
\def\szchit{\mbox{$\sigma_0^{\tiny\chit}$}}
\def\szchij{\mbox{$\sigma_0^{\tiny\chij}$}}
\def\dschiz  {\mbox{$d\sigma^{\chiz} $}}
\def\dschio  {\mbox{$d\sigma^{\tiny\chio} $}}
\def\dschit  {\mbox{$d\sigma^{\tiny\chit} $}}
\def\dschij  {\mbox{$d\sigma^{\tiny\chij} $}}

\def\vd{\mbox{${\bf D}$}}
\def\ve{\mbox{${\bf E}$}}
\def\vb{\mbox{${\bf B}$}}
\def\vs{\mbox{${\bf \sigma}$}}

\def\rj {{\mathrm J}}

\def \chizgg {\mbox{$\Gamma(\chiz \to \gamma\gamma)$}}
\def \chitgg {\mbox{$\Gamma(\chit \to \gamma\gamma)$}}
\def \chijgg {\mbox{$\Gamma(\chi_J \to \gamma\gamma)$}}
\def \chizlh {\mbox{$\Gamma(\chiz \to \lh)$}}
\def \chiolh {\mbox{$\Gamma(\chio \to \lh)$}}
\def \chitlh {\mbox{$\Gamma(\chit \to \lh)$}}
\def \chijlh {\mbox{$\Gamma(\chi_J \to \lh)$}}

\def\aem{\mbox{$\alpha_{{\mathrm\tiny em}}$}}
\def\coll{\mbox{$ \vert\vert$}}
\def\Qb{\mbox{$\overline Q$}}
\def\ko{\mbox{$k_1$}}
\def\kt{\mbox{$k_2$}}
\def\mqs{\mbox{$m_Q^2$}}
\def\ks{\mbox{$k^2$}}
\def\nep{\mbox{$N_\epsilon$}}
\def\mzs{\mbox{$m_0^2$}}
\def\mos{\mbox{$m_1^2$}}
\def\mds{\mbox{$m_2^2$}}
\def\mts{\mbox{$m_3^2$}}
\def\mfs{\mbox{$m_4^2$}}
\def\tu{\mbox{$t_1$}}
\def\uu{\mbox{$u_1$}}
\def\li{\mbox{$\mathrm{Li}_2$}}
\def\meas{\mbox{$\frac{\rd^D k}{(2\pi)^D}$}}
\def\fc{\mbox{${\mathrm F} \chi$}}
\def\nfc{\mbox{${\mathrm NF} \chi$}}

\def\mufrag{\mbox{$\mu_{\mathrm F}$}}
\def\mufrags{\mbox{$\mu^2_{\mathrm F}$}}

\def\mct{\mbox{$m_c^3$}}
\def\mcf{\mbox{$m_c^5$}}

\def\msb{\mbox{${\overline{MS}}$}}
\def\shad{\mbox{$S_{\mathrm had}$}}
\def\nc{\mbox{$N_c$}}
\def\psiq{\mbox{$\psi_Q$}}
\def\pspq{\mbox{$\psi^\prime_Q$}}
\def\chijq{\mbox{$\chi_{QJ}$}}
\def\nj{\mbox{$N_J^\epsilon$}}

\def\bfp{{\bf p}}
\def\bfpp{{\bf p}'}
\def\bfk{{\bf k}}

\def\slash#1{{#1\!\!\!/}}

\def\Da{D_{a\to \gamma}}
\def\Dg{D_{g\to \gamma}}
\def\DBqp{D^B_{q\to \gamma}}
\def\DBq{D^B_{q\to \gamma}}
\def\DBqpp{D^B_{q'\to \gamma}}
\def\DBgp{D^B_{g\to \gamma}}
\def\Dq{D_{q \to \gamma}}
\def\Dqp{D_{q'\to \gamma}}
\def\DBa{D^B_{a\to \gamma}}

\def\Pqqms{{\cal P}_{q\to q}}
\def\Pgqms{{\cal P}_{q\to g}}
\def\Pqgms{{\cal P}_{g\to q}}
\def\Pggms{{\cal P}_{g\to g}}
\def\Ppqms{{\cal P}_{q\to \gamma}}
\def\Ppgms{{\cal P}_{g\to \gamma}}
\def\Pjims{{\cal P}_{i\to j}}

\def\Pgq{P_{q\to g}}
\def\Ppq{P_{q\to \gamma}}
\def\Gij{{\cal G}_{i\to j}}
\def\Gqp{{\cal G}_{q\to \gamma}}
\def\Gqpp{{\cal G}_{q'\to \gamma}}
\def\Ggp{{\cal G}_{g\to \gamma}}
\def\Gqg{{\cal G}_{q\to g}}
\def\Ggq{{\cal G}_{g\to q}}
\def\Gij{{\cal G}_{i\to j}}
\def\Gab{{\cal G}_{a\to b}}
\def\Gqq{{\cal G}_{q\to q}}
\def\Ggg{{\cal G}_{g\to g}}
\def\Gqpq{{\cal G}_{q'\to q}}
\def\Gqqp{{\cal G}_{q\to q'}}
\def\calDq{{\cal D}_{q\to \gamma}}
\def\calDg{{\cal D}_{g\to \gamma}}

\begin{titlepage}
\nopagebreak
{\flushright{
        \begin{minipage}{5cm}
        CERN-TH/98-152\\
        ANL-HEP-PR-98-44\\
        \end{minipage}   }

}
\vfill
\begin{center}
{\LARGE { \bf \sc Colour-octet effects in\\
\vspace{.2cm}
radiative $\Upsilon$ Decays}{\large \footnote{
This work was supported in part by the EU Fourth Framework Programme `Training
and Mobility of Researchers', Network `Quantum Chromodynamics and the Deep
Structure of Elementary Particles', contract FMRX-CT98-0194 (DG 12 - MIHT), and partially by the U. S. Department of Energy, under contract W-31-109-ENG-38.}}}
\vfill                
{\bf Fabio MALTONI\footnote{Permanent address:
     Dipartimento di Fisica dell'Universit\`{a} and Sez. INFN, Pisa, Italy} 
}\\
{CERN, TH Division, Geneva, Switzerland} \\
\verb+fabio.maltoni@cern.ch+\\
\vskip .5cm
{\bf Andrea PETRELLI}\\ 
Argonne National Laboratory, HEP Division\\ 
Argonne, IL, USA\\                     
\verb+petrelli@hep.anl.gov+\\
\end{center}
\nopagebreak
\vfill
\begin{abstract} 
We investigate the effects of colour-octet contributions to the radiative 
\ups decay within the Bodwin, Braaten and Lepage NRQCD factorization 
framework. 
Photons coming both from the coupling to hard processes (`direct')
and by collinear emission from light quarks (`fragmentation')
are consistently included at next-to-leading order (NLO) in $\as$.
An estimate for the non-perturbative matrix elements
which enter in the final result is then obtained. By comparing the NRQCD prediction at NLO for total decay 
rates with the experimental data, it is found
that the non-perturbative parameters must be  smaller 
than expected from the na\"\i ve scaling rules of NRQCD. 
Nevertheless, colour-octet contributions to the shape 
of the photon spectrum turn out to be significant. 
\end{abstract}                                                
\vskip 1cm
CERN-TH/98-152\hfill \\
ANL-HEP-PR-98-44\\ 
\today \hfill
\vfill 
\end{titlepage}

\section{Introduction}
\label{sec:introduction}

Since the early times of QCD, heavy quarkonia decays have been
considered among the most promising processes to test the perturbative
sector of the theory and to extract the value of the strong coupling
at scales of the order of the heavy-quark mass. In addition to the
calculation and comparison of full inclusive decay rates, much
attention has been devoted to the decays in which one photon is
emitted, and its energy measured~\cite{EXP}.  Experimental data on the
direct photon spectrum in $\Upsilon$ decays have been
compared~\cite{Brodsky,Field}, up to now, under the assumption of a
factorization between a short-distance part describing the
annihilation of the heavy-quark pair in a colour-singlet state and a
non-perturbative long-distance factor, related to the value of the
non-relativistic wave function at the origin.

Recently Bodwin, Braaten and Lepage (BBL)~\cite{bbl} provided a new
framework to study quarkonium decay within QCD. Introducing an
effective non-relativistic theory (NRQCD), perturbative and gauge-invariant 
factorization is obtained by including in the decay
intermediate $\QQ$ states with quantum numbers different from those of
the physical quarkonium state. The relative importance of various
contributions depends on short-distance coefficients which are calculable by
standard perturbative techniques, and on long-distance matrix elements,
which can be either extracted phenomenologically from the data or
calculated on the lattice. In the end one is able to
organize all these terms in a double perturbative series in the strong
coupling $\as$ and in the relative velocity $v$ of the heavy quarks,
and then to make predictions at any given order of accuracy.

In quarkonia decays, photons arise from electromagnetic coupling to both
heavy and light quarks. While contributions coming from the former, at
leading order (LO) in $\as$ in the Colour-Singlet Model (CSM)
--i.e. at the lowest order in $v$ expansion in NQRCD-- have been known
for a long time and are one of the first tests of QCD
~\cite{Brodsky,Field}, LO contributions coming from collinear emission
from light quarks have surprisingly  been considered only recently by
Catani and Hautmann ~\cite{Catani}.  The inclusion of these
`fragmentation' contributions within the CSM was found to greatly
affect the photon spectrum in the $\Upsilon$ decay at low values of
the energy fraction taken away by the photon~\cite{Catani}. Moreover,
one finds that at LO such a contribution comes entirely from the gluon,
as the decay into light quarks vanishes.

It then becomes natural to assess to which extent this picture remains
unchanged at next-to-leading order (NLO) in $\as$ and in $v$. The aim of
this work is to investigate the effects of colour-octet intermediate
states on the photon spectrum, at fixed order in $\as,\aem,v$,
including the coupling of the photons to light quarks and gluons. In
fact, while the order of magnitude of octet contributions is predicted
using scaling rules, and found to be suppressed by powers of $v$ with
respect to the LO colour-singlet ones, their short-distance
coefficients receive contributions at lower order of $\as$, and are
then numerically enhanced.  Furthermore, once leading logarithmic
corrections are included, it is found  that, contrary to the
colour-singlet case, quark and gluon fragmentation into a photon
appears at the same order in the $\as, \aem$ expansion and there is 
no signature to distinguish between the two.

The paper is organized as follows. In section~\ref{sec:General} we
summarize the analysis of quarkonium decay into photons
and hadrons in the framework of NRQCD. 
Section~\ref{sec:NLO} describes the NLO
calculation and the technique used to 
isolate and cancel/subtract IR and collinear divergences.
In  section~\ref{sec:Results} we give estimates for the 
non-perturbative matrix elements by comparing the NLO predictions
for total decay rates with experimental data. Finally, we present 
a numerical study of the impact of 
octet states on the shape of the photon spectrum.  
The last section is devoted to our conclusions.
Appendix~\ref{app:A} collects symbols and notation,
appendix~\ref{app:LO} collects the results for the Born decay rates in
$D$ dimensions.
A summary of the NLO results is
provided in appendix~\ref{app:NLO}, where differential decay rates are
presented in their final form,  after  cancellation of all
singularities.

\section{NRQCD and fragmentation}
\label{sec:General}
A consistent description of the  photon energy spectrum in
$\ups\to\gamma\,+X$ decay requires the inclusion of the
fragmentation components~\cite{Catani}.
The differential photon decay can be expressed in
terms of a convolution between partonic kernels $C_a$ and the
fragmentation functions $D_{a\to\gamma}$:   
\ba
\frac{\rd\Gamma}{\rd z} &=& C_\gamma(z) +\sum_{a= q, \overline
q ,g}\int_z^1 \frac{\rd x}{x} C_a(x,\mufrag)
D_{a\to\gamma}(\frac{z}{x},\mufrag) \nn \\
\nn\\
&\equiv& C_\gamma+ \sum_a C_a \otimes \Da \,,
\label{eq:master}
\ea 
where $z= E_\gamma/m_Q$ is the
rescaled energy of the photon ($m_Q$ is the heavy-quark mass).
The first term corresponds to what is usually called
the `prompt' or `direct' photon production 
where the photon is produced directly in the hard interaction 
while the second one  corresponds to  the long-distance
fragmentation process where one of the partons fragments
and transfers a fraction of its  momentum to the photon.

Each type of parton, $a$, contributes according to the process-independent 
parton-to-photon fragmentation functions $\DBa$ and the sum runs over all partons.
Note that although the fragmentation functions are non-perturbative, 
we can assign a power of coupling constants, 
based on naively counting the couplings necessary to radiate a photon:  
since the photon couples directly to the quark, 
$\Dq$ is of ${\cal O}(\aem)$, while 
we might expect that $\Dg$ is of ${\cal O}(\aem \alpha_s)$.
An explicit calculation at leading order in $\as$ gives:
\ba
z \Dq (z) &=& e_q^2 \frac{\aem}{2 \pi}  z \Ppqms (z) \log 
\frac{Q^2}{\Lambda^2}\,,\label{eq:LOfragq}
\\ 
z \Dg (z) &=& 0 \,,
\label{eq:LOfragg}
\ea
where the $\log(Q^2/\Lambda^2)$ in eq.~(\ref{eq:LOfragq}) comes from
the integration over the transverse momentum of the emitted photon and
$\Lambda$ is a collinear cut-off that reveals the breaking of the
perturbative approach and can be chosen of the order of $\Lambda_{\rm
QCD}$.  The photon fragmentation functions evolve with $Q^2$ just as
the usual hadronic fragmentation functions do, as a result of gluon
bremsstrahlung and $\qq$ pair production. Such evolution can be
derived from a set of coupled equations, which are the usual
Altarelli-Parisi equations but with an added term that takes into
account the leading behaviour in eq.~(\ref{eq:LOfragq}). The main
result of the evolution is that $\Dg$ acquires a non-vanishing
contribution so that all the $\Da$ show the typical logarithmic growth
of eq.~(\ref{eq:LOfragq}). This leads to using the following 
leading-log approximation (LLA) for
the fragmentation functions~\cite{Owens}:

\be
  \Da(z,Q) = \frac{1}{b_0} \frac{\aem}{\as(Q)}  f_a(z)\,,
\label{eq:owens}
\ee 
where $f_a(z)$ are to be extracted from the data.
This shows explicitly that in general
the determination of the spectrum at $O(\aem \as^k)$ requires 
the knowledge of partonic kernels $C_a$ in eq.~(\ref{eq:master}) at
$O(\as^{k+1})$.
This observation was first made, in quarkonia decays, by
Catani and Hautmann~\cite{Catani} who evaluated the effects of
fragmentation contributions to the photon energy spectrum within the CSM.  
They found a strong enhancement in the
region of small $z$, where soft radiation becomes dominant.

In the NRQCD perspective, a heavy-quarkonium state is represented by
a superposition of infinite $\QQ$ pair configurations organized in
powers of $v$; ~$v\equiv\langle \stackrel{\to}{v}^2\rangle^{1/2}$
is the average velocity of the heavy quark in the quarkonium rest
frame. Within this framework, the decay width is expanded in terms
of the matrix elements of 4-fermion operators (that create and annihilate a
given $\QQ$ pair) times perturbative coefficients associated to each
operator.  By implementing the NRQCD factorization formalism within
the fragmentation picture, the effects of higher Fock components in the
quarkonium state can therefore be evaluated systematically.

The NRQCD expansion for the coefficients $C_i(x)$ reads: 
\ba
C_i &=& \sum_\q C_i[\q]
\;\;\;\; i=\gamma, q, {\overline q} ,g\; ,
\label{cexpo}\\ 
C_i[\q]&=&\hat
C_i[\q](\as(m_Q),\mul)\; \frac{\langle \ups\vert {\cal O}
(\q,\mul)\vert \ups\rangle}{m^{\delta_\q}}\, , \label{cexpt} 
\ea 
where $\mul$ is the NRQCD factorization scale and
$\hat C_i[\q](x,\as(m_Q),\mul)$  the perturbative coefficients
(here we have dropped the dependence
of $\hat C_i$ on the fragmentation scale $\mufrag$). The NRQCD
sum is performed over all the relevant spin, angular momentum and colour configurations
$\q$ that contribute at a given order in $v$. In the case of a  $\Upsilon$, the structure of the Fock state at order $v^4$ is 
\ba \vert\Upsilon\rangle
= O(1)\vert b\overline b[\psis]\rangle + \sum_J O(v)\vert
b\overline b[\chijh]\rangle + O(v^2)\vert b\overline
b[\etah]\rangle + O(v^2)\vert b\overline b[^3S_1^{[1,8]}]\,\rangle\,. 
\ea 
As a consequence, eq.~(\ref{cexpo}) can be written in the following explicit form:
\ba C_i &=& \hat
C_i[\psis] \frac{\langle\Upsilon\vert{\cal
O}_1(\threeSone)\vert\Upsilon\rangle}{m^2} + \hat
C^\prime_i[\psis] \frac{\langle\Upsilon\vert{\cal
P}_1(\threeSone)\vert\Upsilon\rangle}{m^4} +\sum_J \hat
C_i[{^3P_J^{[8]}}] \frac{\langle\Upsilon\vert{\cal
O}_8(^3P_J)\vert\Upsilon\rangle}{m^4}\nn\\
&+& \hat C_i[\etah]
\frac{\langle\Upsilon\vert{\cal
O}_8(\oneSzero)\vert\Upsilon\rangle}{m^2} + \hat C_i[\psih]
\frac{\langle\Upsilon\vert{\cal
O}_8(\threeSone)\vert\Upsilon\rangle}{m^2}+O(v^6)\,.
\ea 
Let us consider the direct contributions ($i=\gamma$). The leading
colour-singlet dimension-6 operator contribution is of $O(\assq\aem)$, 
and the ${\cal P}_1-{\rm operator}$
contribution is suppressed by $v^2$. All the colour-octet processes
start contributing at $O(\as\aem v^4)$. By naive power counting, and
using the approximate relation $\as \sim v^2$, one finds
therefore that the octet states contribute to the same order as the
singlet relativistic corrections and might be comparable in
size to these. Moreover differential quantities are obviously sensitive to the
details of the kinematics and so it may happen that contributions
that are suppressed by standard counting rules are actually leading, 
in some particular region of the phase space.
\begin{figure}
\begin{center}
{\unitlength1cm
\begin{picture}(12,4)
\epsfig{file=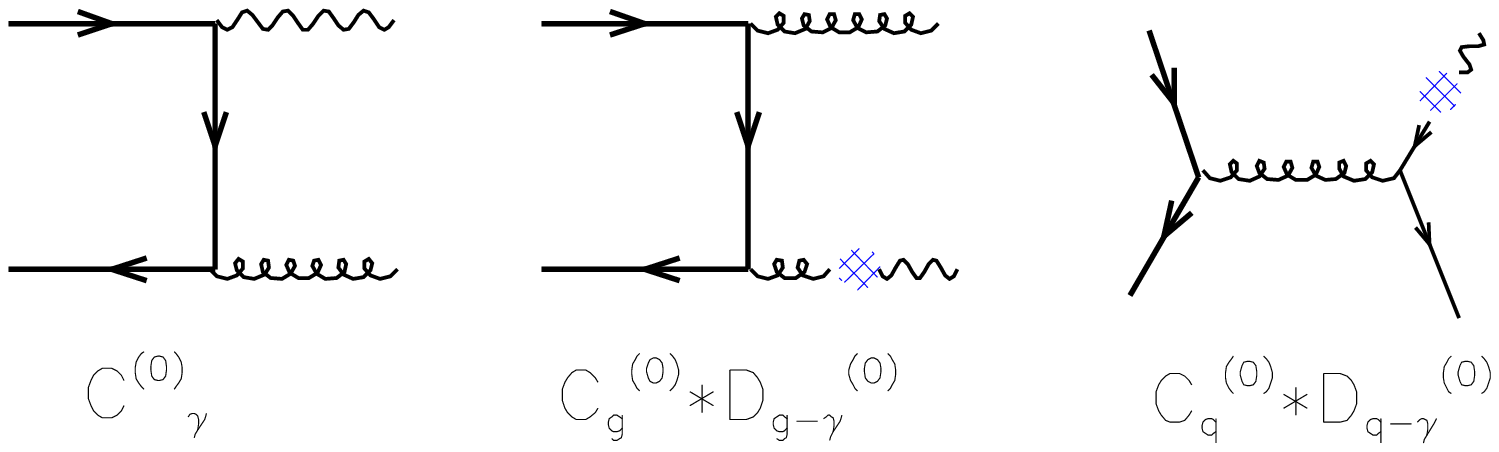,%
          height=4cm,clip=,angle=0}
\end{picture}}
\ccaption{}{\label{fig:diagLO}
Sample of LO Feynman diagrams: direct and  fragmentation.}
\end{center}
\end{figure}

LO diagrams are shown in fig.~\ref{fig:diagLO}.  By considering the
following perturbative QCD expansions of the coefficients $C_a[\q]$
and of the fragmentation functions $D_{i\to j}$,

\ba
C_a[\q] &=& \left(\hat C^{(0)}_a[\q] + \hat
C^{(1)}_a[\q]\right)\frac{\langle\ups\vert{\cal
O}(\q)\vert\ups\rangle}{m^{\delta_\q}}\equiv C^{(0)}_a[\q] +
C^{(1)}_a[\q]+\cdots\, ,\\ D_{i\to j} &=& D^{(0)}_{i\to j} + D^{(1)}_{i\to
j} +\cdots \,,
\ea
one is able to write the general structure of the LO spectrum:
\be
\frac{\rd\Gamma^{(0)}}{\rd z}= \sum_\q \left\{C^{(0)}_{\gamma}[\q] +
 C^{(0)}_{g}[\q]\, \otimes\, \Dg^{(0)} + 2 \,C^{(0)}_{q}[\q]\, \otimes
 \, \Dq^{(0)}\right\}\label{lostru}\,.
\ee
Since the LO colour-octet contributions have a two particle final state, the kinematics is fixed and the delta function $\delta(1-x)$ of the short-distance
coefficient transforms the convolutions in trivial products:
\ba
\frac{\rd\Gamma^{(0)}}{\rd z} &=& \sum_{\q}\left[\Gamma_{\rm Born}(\q\to g\gamma)\delta(1-z)+ 2 \ \Gamma_{\rm Born}(\q\to gg) \Dg^{(0)}(z)\right] 
\nn\\
&&+ 2\,\sum_{q} \Gamma_{\rm Born}(\psih \to \qq)\Dq^{(0)}(z)\, ,
\label{eq:lowidth}
\ea
where the first sum is performed over the lowest-order
non-zero octet configurations 
$\q =\oneSzero^{[8]}$, $\threePzero^{[8]}$, $\threePtwo^{[8]}$, while 
the second one over the flavours of the light quarks.
As eq.~(\ref{eq:lowidth}) shows, at leading order the colour-octet contributions are proportional to the fragmentation functions and to terms
proportional to $\delta(1-z)$ which do not contribute
for $z<1$\footnote{Although we did not include these `direct' terms
in our analysis, we expect that resummation of higher order effects 
for $z \sim 1$ will induce an effective smearing of the delta function and 
'feed down' some photons to lower values of $z$~\cite{RW}. 
This point will be discussed in more detail in the sequel.}. 

\begin{figure}[t]
\begin{center}
{\unitlength1cm
\begin{picture}(12,10)
\psfig{file=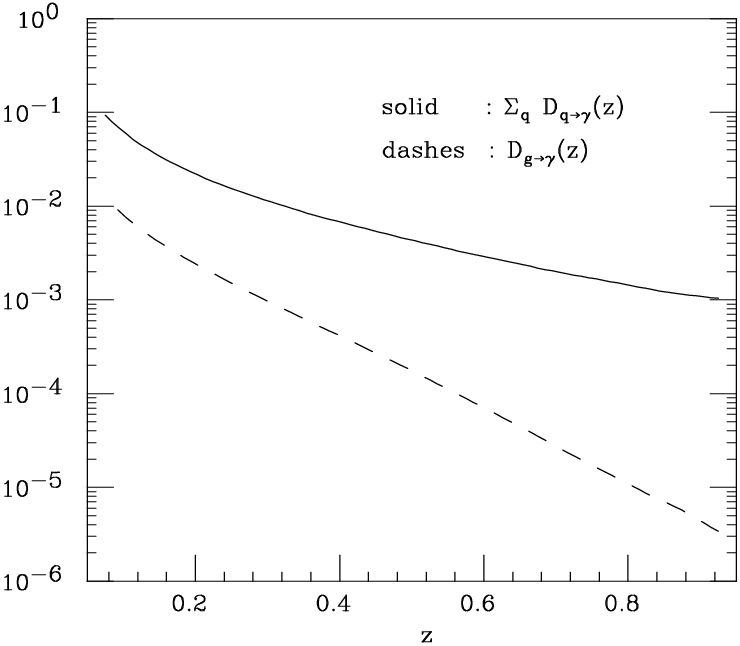,height=9cm,clip=,angle=0}
\end{picture}}
\ccaption{}{\label{fig:plot-FF}
Fragmentation functions of a light parton into a photon according to the reference~\cite{fragm}.}
\end{center}
\end{figure}

The fragmentation functions of a light parton
into a photon have been calculated by several groups
\cite{fragm,ffgerman}. In this paper we employ the set recently
developed by Bourhis, Fontannaz and Guillet \cite{fragm}.
In fig.~\ref{fig:plot-FF} functions $D_{g \to \gamma}(z)$ and
${\sum}_q D_{q \to \gamma}(z)$ are shown: as  was previously stated,
the contribution from quarks is dominant.

\section{NLO radiative decays: the calculation technique}
\label{sec:NLO}
In this section we briefly describe the strategy for the calculation of
higher-order corrections.
A consistent calculation of these 
entails the evaluation of the real and virtual emission diagrams,
carried out in $D$ dimensions. The UV divergences present in the
virtual diagrams are removed by the standard renormalization. The IR
divergences appearing after the integration over the phase space of
the emitted parton are cancelled by similar divergences present in the
virtual corrections, or by higher-order corrections to the
long-distance matrix elements~\cite{bbl}.  Collinear divergences,
finally, are either cancelled by similar divergences in the virtual
corrections or by factorization into the NLO fragmentation functions.  
The evaluation of the real emission matrix elements in $D$ dimensions
being particularly complex, we follow in this paper 
the technique developed in  ref.~\cite{Mele91} and already
employed in~\cite{Petrelli97,Maltoni97}, 
whereby the structure of soft and collinear
singularities in $D$ dimensions is extracted by using universal
factorization properties of the amplitudes. Thanks to these
factorization properties, the residues of all IR and collinear poles
in $D$ dimensions can be obtained without an explicit calculation of
the full $D$-dimensional real matrix elements. In general they only require
the knowledge of the $D$-dimensional Born-level amplitudes, a
much simpler task.  The isolation of these residues allows  
the complete cancellations of the relative poles in $D$
dimensions to be carried out, 
leaving residual finite expressions, which can then be
evaluated exactly directly in $D=4$ dimensions. In this way one can
avoid the calculation of the full $D$-dimensional real-emission matrix
elements.  Furthermore, the four-dimensional real matrix elements that
will be required have been known in the literature for quite some
time~\cite{Cho96,Cacciari96}.
The study of the soft behaviour of the real-emission amplitudes
was already presented in ~\cite{Petrelli97,Maltoni97} and 
we made substantial use of those results.

To be more specific, let us consider the three-body decay processes 
$\q^{[1,8]} \to k_1 + k_2 + k_3$, 
where $\q^{[1,8]}\equiv\QQ[\spectr^{[1,8]}]$. 
Using the conservation of energy-momentum and rotational invariance,
it is straightforward to verify that there are only two independent
variables, which we chose to be $x_i$, the fraction of energy of
the parton whose spectrum we are interested in, and $y$ ,
the cosine of the angle of such parton with one of the other two.
Within this choice, the differential decay width in $D$ dimensions 
reads :
\ba                                    
\label{eq:gamma}
  C_i^{(1)}[\q]&=& 
\C \, \frac{1}{{\cal S}_1}\;
         x_i^{1-2\eps} \, (1-x_i)^{-1-\eps} \, \int_0^1 \rd y \left[ y(1-y) \right]^{-1-\eps}\,f_R[\q](x_i,y)\nn\\                    
&&+\frac{\Phi_{(2)}}{2M {\cal S}_2} f_V[\q]\,\delta(1-x_i)\; \equiv \;C_i^{(R)}[\q] + C_i^{(V)}[\q]\ . 
\label{eq:diffdecay}
\ea               

\begin{figure}
\begin{center}
{\unitlength1cm
\begin{picture}(14,10)
\epsfig{file=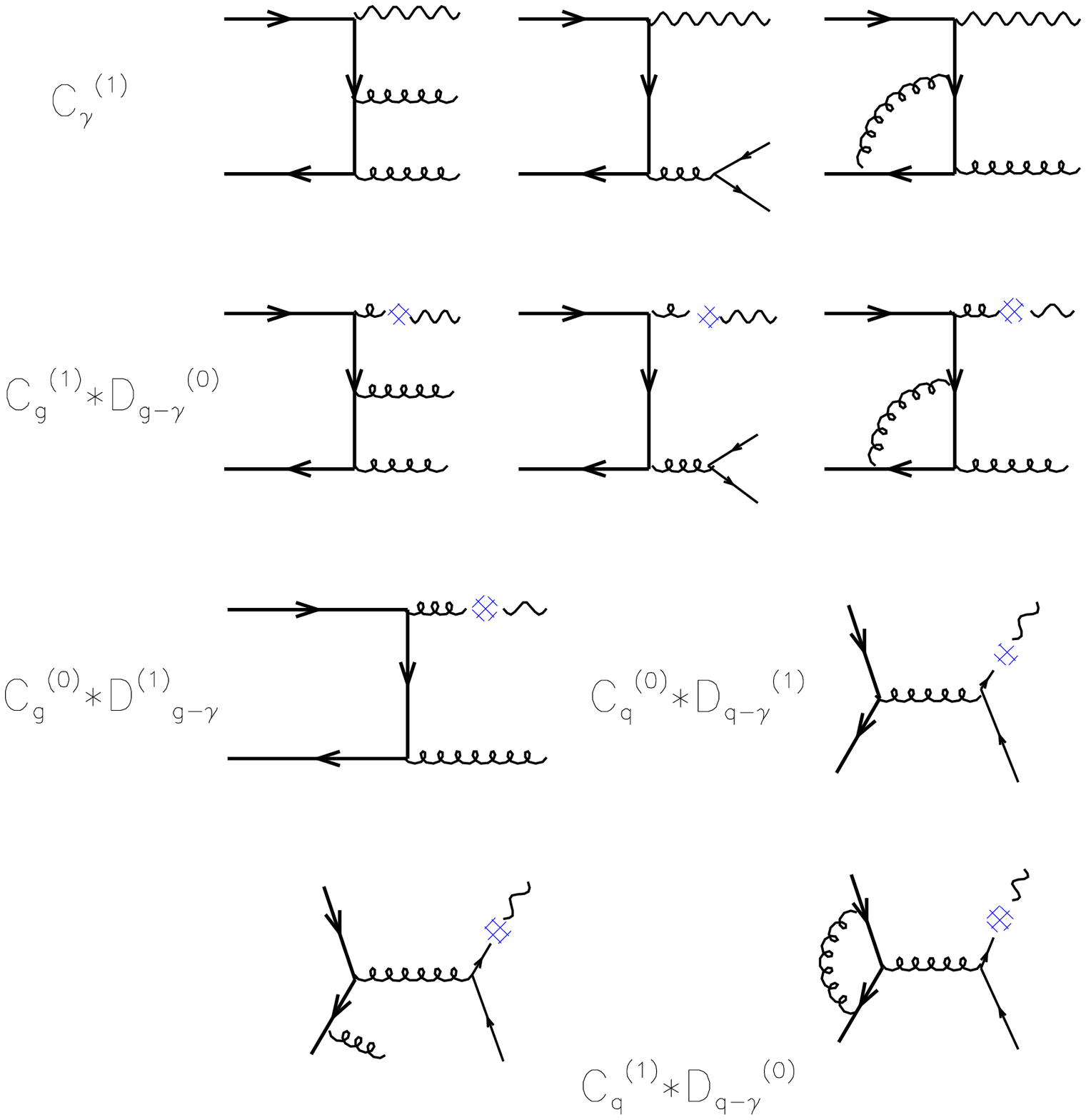,%
          height=12cm,clip=,angle=0}
\end{picture}}
\ccaption{}{\label{fig:diagNLO}
Sample of NLO Feynman diagrams: direct and fragmentation.}
\end{center}
\end{figure}   
The NLO spectrum coefficients are the sum of the virtual and the real~($R$) and the virtual~($V$) QCD corrections. In general  both channels $ggg$ and $\qq g$ contribute to the real term, the  
${\cal S}_{1,2}$  are factors that account for the right
counting for identical particles in the final state, 
and for the multiplicity of the various corrections, and
$\Phi_{(2)}$ is the total two-body phase space in $D$ dimensions:
\be                                                                     
      \Phi_{(2)} \= \frac{1}{8\pi }
       \left(\frac{4\pi}{M^2}\right)^{\eps}
      \; \frac{\Gamma(1-\eps)}{\Gamma(2-2\eps)} \; ,
\label{eq:phit}
\ee 
while $N$ and $K$ are defined as
\ba         
    N = \frac{M^2}{(4\pi)^2} \; \left( \frac{4\pi}{M^2} \right) ^{\eps}
      \; \Gamma(1+\eps) \; ,                                             
    &\;\;\;& K = \Gamma(1+\eps)\Gamma(1-\eps) \sim 1+\eps^2\frac{\pi^2}{6} \;.
\ea
The function $f(x,y)$ is defined as 
\ba
f_R[\q](x_i,y) &=&  (1-x_i) y (1-y) \overline{\sum}\left\vert A_R[\q](x_i,y)\right\vert^2\\ \,
f_V[\q] &=& 2\, {\rm Re} \overline{\sum}\left(A_B A_V^*\right) \,.
\label{eq:nlocalc}
\ea 

Since divergences can appear only at the border of phase space,
i.e. $y=0, y=1, x_i=0, x_i=1$, $f_R$ is finite for all values of $x$
and $y$ within the integration domain.  Therefore all singularities of
the total decay rates can be easily extracted by isolating the
$\eps\to 0$ poles from those factors in eq.~(\ref{eq:gamma}) that 
explicitly depend on $x_i$ and $y$.  It must be noted that an infrared
divergence arises in the limit $x_i\to 0$ when $i=g$, giving a term of
the form $\sim\log x_g$ in the width.  Nevertheless we are not
interested in regularizing such a divergence, since, in this case, the
physical resolution of the detector works as a physical cut-off. For
the same reason the virtual gluon emission at $x_g=0$ has not been
included in the account of the multiplicities.
 
The virtual coefficients can be extracted straightforwardly from 
ref.~\cite{Maltoni97}. The calculation of the real coefficients is
much more complicated, and it has been carried out by exploiting the
soft properties of the amplitude obtained in refs. 
~\cite{Petrelli97,Maltoni97}.  To illustrate the fundamental
steps of the calculation of the real part, we consider here
the $C^{(R)}_g[\q]$ coefficients, with $\q$ being one among the $C$-even 
configurations
$\etah,\chizh,\chith$.  Let also $\nf=0$ for the time being, so that we
neglect contributions coming from the decay into $\qq g$.  In this
case we reorganize the first term of eq.~(\ref{eq:nlocalc}) by
expanding the structure in powers of $\epsilon$ and using the symmetry
of the phase space. Considering the spectrum of the gluon ``1'', we
find
\ba
C_g^{(R)}[\q] &=&\C\,\frac{1}{{\cal S}_1} \left[ 2\, \left( \frac{1}{1-x}\right)_+ f_R[\q](x,0)\left(-\frac{1}{{\epsilon}_{\rm coll}} + 2\,\log x\right) \right.\nn\\
&+& \left.2 x\,\left( \frac{\log(1-x)}{1-x}\right)_+ f_R[\q](x,0) -\frac{1}{\epsilon}\delta(1-x)\int^1_0\rd y [y(1-y)]^{-1-\epsilon} f_R[\q](x,y)
\right.\nn\\
&+& \left.2 \left( \frac{1}{1-x}\right)_+\int_0^1 \rd y \left(\frac{1}{y}\right)_+  f_R[\q](x,y) \right]\, .
\label{eq:realexp}
\ea
\begin{figure}[t]
\begin{center}
{\unitlength1cm
\begin{picture}(12,10)
\psfig{file=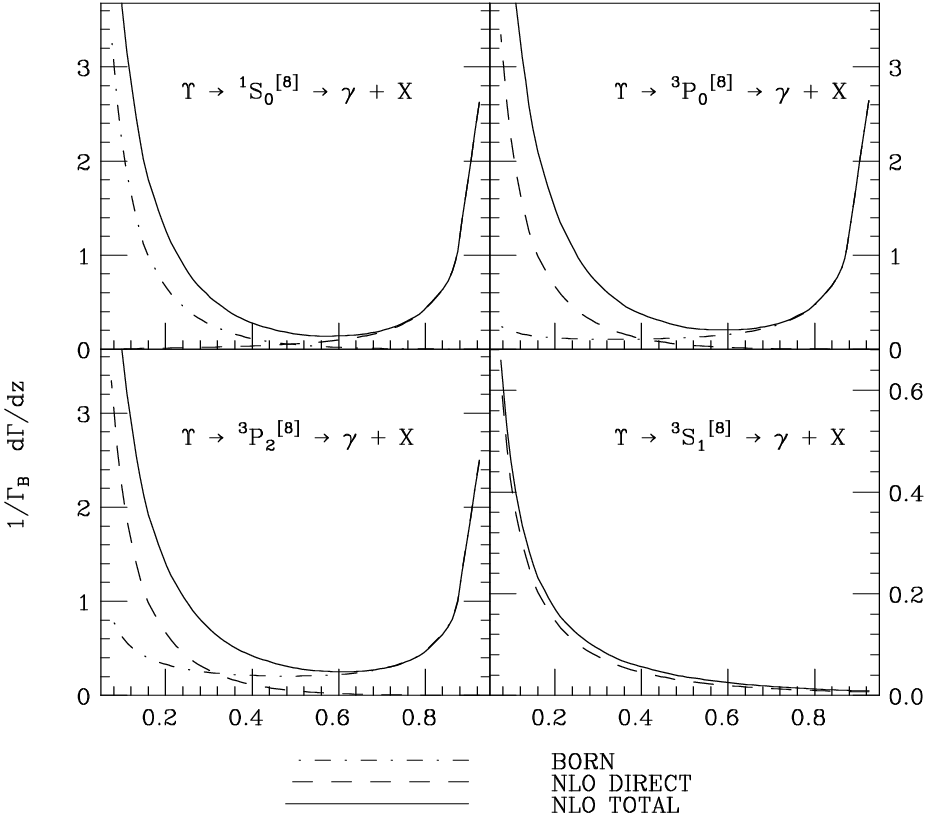,%
          height=10cm,clip=,angle=0}
\end{picture}}
\ccaption{}{\label{OCT}
Different colour-octet contributions to the photon spectrum in the $\Upsilon$ decay up to $O(v^4)$. The differential decay widths $\rd \Gamma/\rd z$ are reported as a function of $z=E_\gamma/m$.
All the distributions displayed are normalized to the respective Born (that is $\q\to g\gamma$ for $C$-even and $\psih\to\qq$ for the only $C$-odd contribution).}
\end{center}
\end{figure}

The soft divergences $\sim \delta(1-x)$ cancel by adding the virtual
contribution in the same area of the phase space.  The last piece of
eq.~(\ref{eq:realexp}) is a state-dependent finite
contribution. The limit $y\to 0$ corresponds to gluon 1 and gluon 2
becoming collinear $1|\!|2$ and the factor 2 in front accounts for the
case $1|\!|3$.  Integration over the phase space gives rise to a pole
labelled by $\epsilon_{\rm coll}$ and a universal finite part.  This
divergence is not cancelled by adding the virtual term and reveals
that non-perturbative effects are leading in this case.  In fact the
residual sensitivity can be consistently factorized  into the
fragmentation function of the gluon into the photon.  Such singular
residual collinear part corresponds to the first term in 
eq.~(\ref{eq:realexp}) plus the collinear piece of the virtual 
contribution ($\sim \delta(1-x)$) 
that comes from the gluon, ghost self-energy loops of
the gluon we are selecting, so that it reads

\ba
C_g^{({\rm coll})}[\q] &=& -\frac{1}{\epsilon_{\rm coll}}\left(\frac{4 \pi\mu^2}{M^2}\right)^\epsilon\Gamma(1+\epsilon)\frac{\as}{\pi}\times\nn\\
&&\left\{2\,\ca\left[\frac{x}{(1-x)_{+}}+\frac{1-x}{x}+x(1-x)\right] + \frac{11}{6}\ca\delta(1-x)\right\}\Gamma_{\rm Born}[\q]\,.
\ea
If we now switch on the light flavours including $\qq g$ and gluon vacuum polarization diagrams, then we obtain the conventional counter-term $\sim{\cal P}_{gg}$, which has to be subtracted at the factorization scale $\muf$.

This procedure can be extended to all short-distance terms and may be useful
to express the factorization in a more general way. 
At NLO the individual terms in eq.~(\ref{eq:master})
may be divergent and will be denoted by tilded quantities.
As we have already mentioned,
such divergences correspond only to two final partons
becoming collinear, and their form is dictated
by the factorization theorem. 
According to this
we can reorganize them as follows:
\ba
\tilde C_\gamma  &=&   C_\gamma 
+ \sum_q   C_q \otimes \Gqp
+   C_g \otimes \Ggp, \nonumber \\
 \tilde C_q  &=& \sum_{q'}   C_{q'}\otimes \Gqpq 
+   C_g\otimes \Ggq, \nonumber \\
 \tilde C_g  &=& \sum_{q}   C_{q}\otimes \Gqg + 
  C_g\otimes \Ggg, 
\label{eq:gstruc4}
\ea
where all of the divergences are now concentrated  in the 
factorization-scale-dependent, transition functions $\Gij$:
\ba
\Gab &=& \;\delta_{ab}\, \delta(1-x)\, +
\,\frac{\as }{2\pi} 
\frac{1}{\Gamma(1-\epsilon)}
\left(\frac{4\pi\mu^2}{\mu_{F}^2}\right)
\left[-\frac{1}{\epsilon}{\cal P}_{ba}(x)\right] +K_{ab}(x)\,,
\\
\Ggp &=& K_{g\gamma}(x)\,,\\
\Gqp &=& \left(\frac{\aem  e_q^2}{2\pi}\right)
\frac{1}{\Gamma(1-\epsilon)}\left(\frac{4\pi\mu^2}{\mu_{F}^2}\right)
^{\epsilon}\,
\left[-\frac{1}{\epsilon}{\cal P}_{\gamma q}\right]+K_{q\gamma}(x)\,, 
\ea
where $a=g,q,\overline q$ and  all the coefficients $C_i$ 
are now finite for $\eps \to 0$.  
The functions ${\cal P}_{ba}(x)$ are the $D=4$  Altarelli-Parisi splitting
kernels, collected in appendix~\ref{app:A}, and the factors $K_{ij}$ are
arbitrary functions, defining the factorization scheme. In this paper we adopt
the $\MSB$ factorization, in which $K_{ij}(x)=0$ for all $i,j$. 
The collinear factors $\Gij$ are usually absorbed into 
the bare fragmentation functions by defining
\ba
\Dq & = & \Gqp + \sum_{q'} \Gqqp \otimes \DBqpp + \Gqg \otimes \DBgp \,,
\nonumber\\
\Dg & = & \Ggp + \sum_{q} \Ggq \otimes \DBqp + \Ggg \otimes \DBgp \,,
\ea
so that we can write the physical decay rate in terms of finite
quantities,
\ba
\frac{{\rm d}\Gamma}{\rd z}(\gamma + X)=C_\gamma+ \sum_q 
C_q \otimes \Dq + C_g \otimes \Dg\, .
\ea

As illustrated in  fig.~\ref{fig:diagNLO}, 
we write the general structure of the NLO processes as
\ba
\frac{\rd\Gamma^{(1)}}{\rd z}=\sum_\q \left[C^{(1)}_{\gamma}[\q] \right.&+&\left.  
 C^{(1)}_{g}[\q] \,\otimes \,\Dg^{(0)} +  C^{(0)}_{g}[\q]\,\otimes\,\Dg^{(1)}\nn\right.\\ &+& \left. 2 \,C^{(1)}_{q}[\q]\,\otimes\,\Dq^{(0)}+ 2 \, C^{(0)}_{q}\,\otimes\,\Dq^{(1)}\right] \,. 
\label{nlostru}
\ea

\section{Results}
\label{sec:Results}
In the previous section we have shown how 
the short-distance coefficients have been calculated and 
all the final-state collinear divergences have been consistently absorbed
into fragmentation functions.  
Now, in order to investigate  the phenomenological applications
of colour-octet states, an estimate of the NRQCD matrix elements (ME)
must be given.
The long-distance MEs can be calculated on the lattice, 
extracted from experiments when enough data are available, 
or roughly determined by using
scaling rules of NRQCD or by renormalization group (RG) arguments.
At the present time  none of the aforementioned 
techniques is able to give a set of precise values for MEs and,
as we will see below, estimates are affected by large uncertainties.
\begin{figure}[t]
\begin{center}
{\unitlength1cm
\begin{picture}(12,10)
\psfig{file=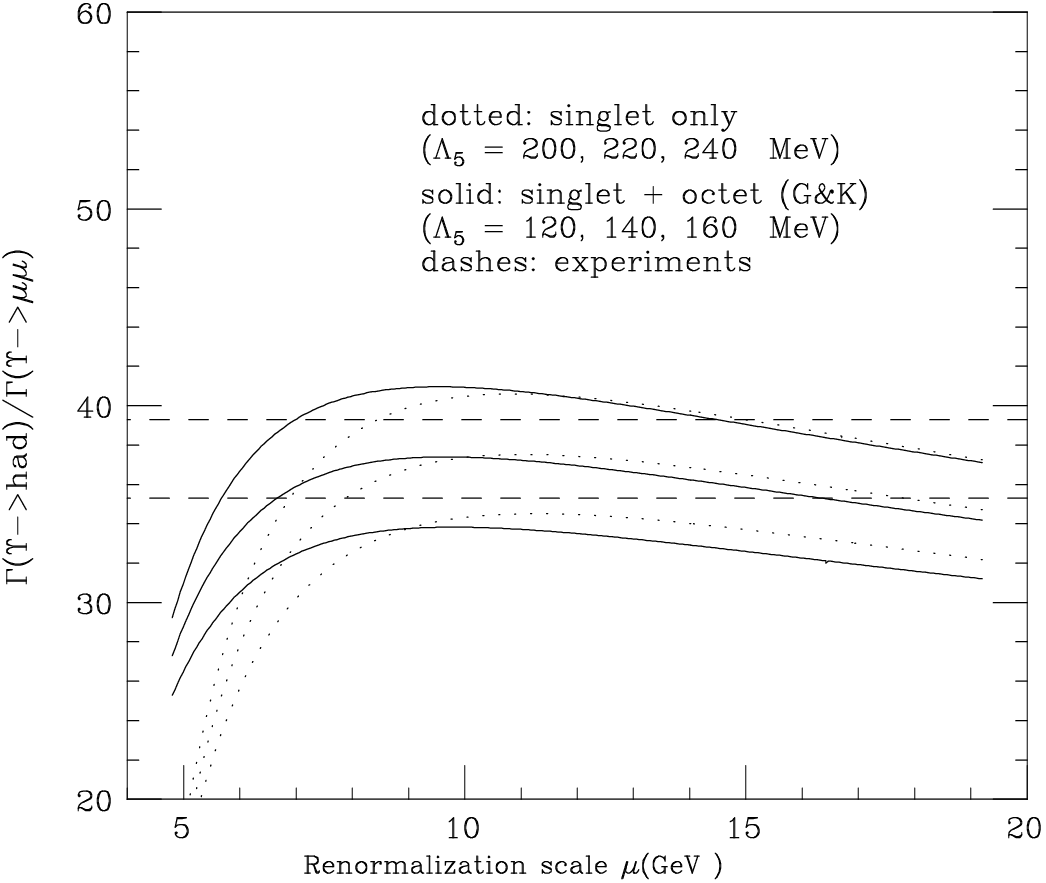,
          width=10cm,clip=,angle=90}
\end{picture}}
\ccaption{}{\label{mes1}
 Ratio $\Gamma(\ups\to{\rm had})/\Gamma(\ups\to\mu^+\mu^-)$ versus
 renormalization scale $\mu$ for different values of $\Lambda_5$. The
 solid lines include NLO colour-octet contributions with the RG
 estimates of the matrix elements. The dotted lines include
 colour-singlet only.}
\end{center}
\end{figure}

\begin{figure}[t]
\begin{center}
{\unitlength1cm
\begin{picture}(12,10)
\psfig{file=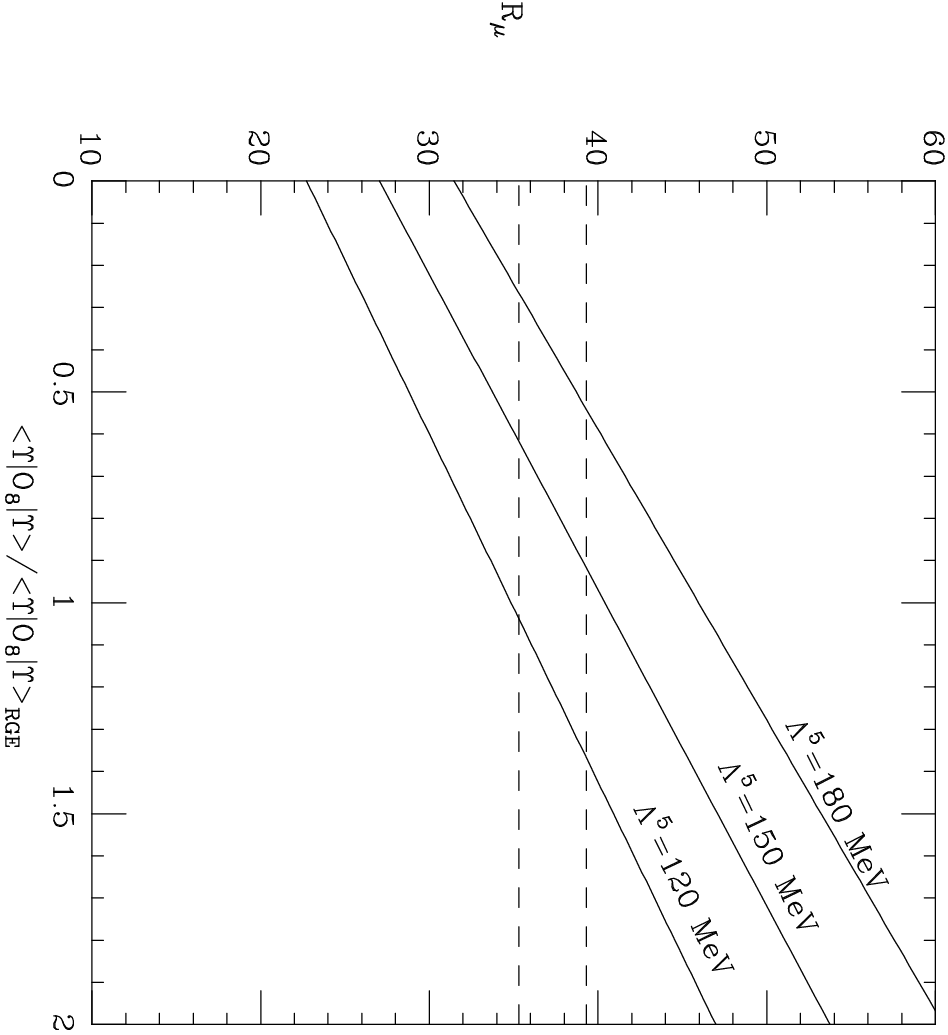,%
          width=10cm,clip=,angle=90}
\end{picture}}
\ccaption{}{\label{mes2} 
Ratio $\Gamma(\ups\to{\rm had})/\Gamma(\ups\to\mu^+\mu^-)$ versus colour-octet
matrix elements for different values of $\Lambda_5$. 
The dashed lines indicate the $2\sigma$ interval of the experimental value
for $R_\mu$.} 

\end{center}
\end{figure}

The velocity-scaling of the MEs is basically determined by the
number of derivatives in the respective operators and by the number
of electric or magnetic dipole transitions between the $\QQ$ pair 
annihilated at short distance and the $\QQ$ pair in the 
asymptotic physical state.
This can nicely be described by a multipole expansion of the
non-perturbative transition $\Upsilon\, \to\, {\cal Q}$:
an  $\etah$ can be reached by a chromo-magnetic dipole transition,
an $\psih$ by a double chromo-electric emission and 
 $\chijh$  by a simple chromo-electric transition.
The first two are of order $v^4$ while the last only of order
$v^2$. Finally, since the hard-production vertex for a $P$-wave
is already suppressed by $v^2$ relative to the production
of an $S$-state, one realizes that the colour-octet $C$-even states 
and $\etah$ all contribute at the same order in $v$. 
Following this approach, we can write:  
\ba
\bra{\Upsilon}{\cal O}_8(\threeSone)\ket{\Upsilon} \approx\;v^4 \;\bra{\Upsilon}{\cal O}_1(\threeSone)\ket{\Upsilon}&\;\;\;& \bra{\Upsilon}{\cal O}_8(\threePJ)\ket{\Upsilon} \approx\;m^2 v^4 \;\bra{\Upsilon}{\cal O}_1(\threeSone)\ket{\Upsilon}\, ,
\label{MEI}
\ea
where for bottomonium one usually takes  $v^2 \simeq 0.1$ 
and $m\simeq 4.8$ GeV. 

\begin{figure}[t]
\begin{center}
{\unitlength1cm
\begin{picture}(12,10)
\epsfig{file=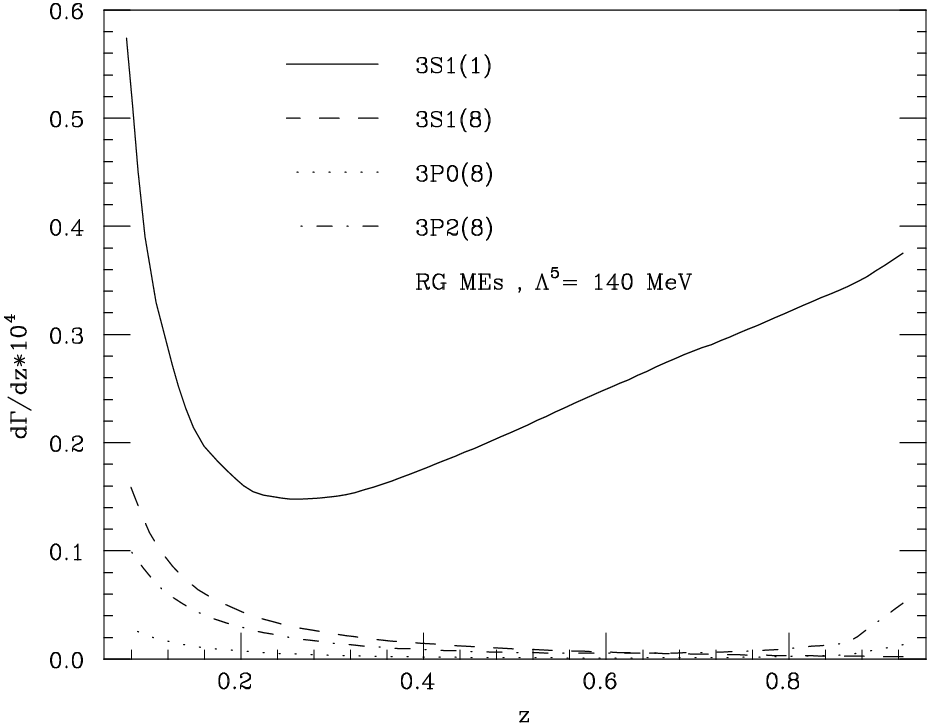,%
          height=10cm,clip=,angle=0}
\end{picture}}
\ccaption{}{\label{fig:plot-res1}
Various Fock contributions to the photon spectrum as a function of $z=E_\gamma/m$. The solid line gives the LO singlet contribution. 
Fragmentation and NLO direct are summed up for each colour-octet state.
The NRQCD MEs are related to the colour-singlet one through the RG estimate. The colour-singlet matrix element is 
arbitrarily chosen to be $\langle\Upsilon|\oo_1(\threeSone)|\Upsilon\rangle = 
 M^2/4\pi$, so that comparison with ref. \cite{Catani} is straightforward.}
\end{center}
\end{figure}

An alternative approach has been considered  
by Gremm and Kapustin in~\cite{GK}. They
obtain estimates  for the colour-octet operators by solving the
RG equations. To order $v^4$ and leading order in $\as$, 
they read:
\ba
\Lambda \frac{d}{d\Lambda}
\bra{\Upsilon}{\cal O}_8 (\oneSzero) \ket{\Upsilon} &=& O(\as v^6),\\
\Lambda \frac{d}{d\Lambda}
\bra{\Upsilon}{\cal O}_8 (\threeSone) \ket{\Upsilon} &=&
\frac{24 \Bf \as}{ \pi m^2}
\bra{\Upsilon} {\cal O}_8 (\threePzero) \ket{\Upsilon},\label{diff}\\
\Lambda \frac{d}{d\Lambda}\bra{\Upsilon}
{\cal O}_8(\threePzero)\ket{\Upsilon} &=&
\frac{8 C_F\as}{81 \pi}  (m v^2)^2
\bra{\Upsilon}{\cal O}_1(\threeSone)\ket{\Upsilon}\,,\label{diff2}
\ea
where we used the heavy quark spin symmetry to reexpress the expectation
values of ${\cal O}_8(^3P_{1,2})$ in terms of ${\cal O}_8(^3P_{0})$.
We note here that our normalization for the colour-singlet NRQCD
operators differs from the original one introduced by BBL, i.e. ${\oo
}_1 = {1\over{2 N_c}} {\o }_1^{\rm BBL}$. Equation ~(\ref{diff}) differs
from the respective equations that appear in  ref.~\cite{GK} because
we included  the contribution of $\chioh$ to the
evolution of ${\cal O}_8 (\threeSone)$, which was left
out in the previous treatment~\footnote{ The authors of 
ref.~\cite{GK} agree that it is correct to include the $\chioh$
contribution in the right-hand side of  eq. (\ref{diff}) (private communication).}.  Assuming that logarithmic terms of the evolution are dominant~\cite{bbl,GK} over the MEs evaluated at a starting scale $\Lambda \sim \Lambda_{QCD}$, we obtain:

\begin{eqnarray}
\bra{\Upsilon}{\cal O}_8(\threeSone)\ket{\Upsilon}_{\rm RG}
&\approx&
 \frac{32\Bf \cf}{27} v^4
 \left(\frac{1}{\b0}\log\left(\frac{1}{\alpha_s(m)}\right)\right)^2
 \bra{\Upsilon}{\cal O}_1(\threeSone)\ket{\Upsilon}\,,\label{MEIIa} \\
\bra{\Upsilon}{\cal O}_8(\threePzero)\ket{\Upsilon}_ {\rm RG}
&\approx&
 \frac{8 C_F}{81 } m^2  v^4 \,\frac{1}{\b0}
 \log\left(\frac{1}{\alpha_s(m)}\right)
 \bra{\Upsilon}{\cal O}_1(\threeSone)\ket{\Upsilon}\,,\label{MEIIb}\\
\bra{\Upsilon}{\cal O}_8(\oneSzero)\ket{\Upsilon}_{\rm RG}&\approx& 0\,.
\label{MEIIc}
\ea

Once numbers are plugged into the previous expression, one realizes
that MEs in eqs.~(\ref{MEI}) result larger by more than one order of
magnitude with respect to the RG estimates shown in
eqs.~(\ref{MEIIa})-(\ref{MEIIc}).  This suggests that the very
first assumption, i.e. that the non-perturbative matrix elements
should be dominated by QCD evolution, is doubtful and cannot be
justified unless their input values were accidentally much smaller than
the `natural' values given in eqs.~(\ref{MEI}). In any case, therefore,
the estimates in eqs.~(\ref{MEIIa})-(\ref{MEIIc}) provide a lower limit for 
the range of all possible values.

To obtain an independent test of the RG estimates and possibly find an
upper limit for the matrix elements, we have analysed their impact on
the observable $R_{\mu}(\Upsilon)=\Gamma(\ups\to{\rm
had})/\Gamma(\ups\to\mu^+\mu^-)$.  This observable is particularly
advantageous because of the cancellation of several sources of
uncertainties: both the colour-singlet NRQCD matrix element and
the overall dependence on the bottom mass cancel in the ratio.
As a result the mass enters only in the logarithm of the renormalization
scale and its  uncertainties can be naturally associated to the choice
of the scale itself.

In fig.~\ref{mes1}, the ratio $R_\mu$ is plotted versus the
renormalization scale $\mu$ (i.e. ~the NRQCD factorization scale is
kept equal to the renormalization one); $\Gamma_{\mu\mu}$ includes the
${\cal P}_1(^3S_1)$-operator contribution~\cite{bbl} and the NLO QCD
corrections~\cite{barbieri}; $\Gamma_{\rm had}$ includes the NLO QCD
colour-singlet~\cite{lepmk} and the ${\cal P}_1(^3S_1)$-operator
contribution~\cite{disq,GK} (dotted curves) and the NLO colour-octet
$\psih$,$\etah$,$\chijh$ \cite{Petrelli97} (solid curves).  The dashes
lines limit the $2\sigma$ band of the experimental value of
$R_{\mu}=37.3\pm 1.0$ \cite{pdg}.  The theoretical curves are drawn
according to the following choice of parameters: $v^2=0.1$ and
$\alpha_{\rm EM}(m_b)=1/132$.  Hence fig.~\ref{mes1} shows that, once
the colour-octet RG MEs estimation is plugged in, the ratio $R_\mu$ is
consistent with the experiments only for $\Lambda_5\simeq 140 \;{\rm
MeV}$ ($\as(M_Z)\simeq 0.110$). On the other hand if we drop the
colour-octet term, just the NLO colour-singlet contribution can still
reproduce the experimental measure of $R_\mu$ by choosing a much
higher value of $\Lambda_5$, namely $\Lambda_5\simeq 220 \;{\rm MeV}$
($\as(M_Z)\simeq 0.118$).

Now we fix the renormalization scale $\mur=10$ GeV. We note that, more than
corresponding to the `natural' choice $\mur \simeq M_{\Upsilon}$, this value
also satisfies the so-called `minimal sensitivity principle'~\cite{Stevenson}, 
i.e. it is the value at which $\mur \frac{d}{d\mur} R_\mu(\mur)$ vanishes. 
Within this choice, we plot the ratio $R_\mu$ versus the variable 
\ba
x = \frac{\langle \Upsilon| \oo_8(\threePzero)|\Upsilon\rangle} {\langle \Upsilon| \oo_8(\threePzero)|\Upsilon\rangle_{\rm RG}} = \frac{\langle \Upsilon| 
\oo_8(\threeSone)|\Upsilon\rangle}{\langle \Upsilon| \oo_8(\threeSone)|\Upsilon\rangle_{\rm RG}}\,.
\ea
The result is shown in fig.~\ref{mes2}. The solid lines represent 
the theoretical calculation of $R_\mu$ and the dashed lines are the $2\sigma$
experimental range, as in fig.~\ref{mes1}.
The larger  the colour-octet 
MEs are, the smaller $\Lambda_5$ has to be taken.
In particular, already for values of the MEs of the order 
of twice the RG estimates, we would find a value of $\Lambda_5 \simeq 80$ MeV 
($\as(M_Z)\simeq 0.102$), well outside the present world average range. 

Following this line one finds that 
the MEs provided by the velocity scaling rules are strictly excluded.
In an ideal global fit perspective both the value of $\Lambda_5$ 
and the colour-octet MEs should be extracted from the data.
Unfortunately the experimental inputs in the $\Upsilon$ decay 
sector are not sufficient to perform a fit of 
such a large number of unknown parameters.

As a confirmation of what we found in fig.~\ref{mes1}, fig.~\ref{mes2}
shows that the RG estimate reproduces the experimental value of $R_\mu$ for $\Lambda_5\simeq 140 \;{\rm MeV}$.
Such a value of $\Lambda_5$  corresponds to $\as(m_b) \simeq 0.190$
and $\as(M_Z) \simeq 0.110$. 
The world average of \as\ ($\as(M_Z)=0.119\pm 0.004$) 
(or equivalently $\Lambda_5\simeq 237$ MeV) is actually consistent with a vanishing (or even negative) octet contribution to the $\Upsilon $ decay into hadrons. Nevertheless the uncertainties involved are still large: 
NNLO QCD corrections (reflected in the $\mu$ dependence of the NLO correction) might be important as well as higher twist effects.
A clear indication that higher order effects are not negligible, comes
from the two-loop calculation of the leptonic width recently performed 
by Beneke et al.~\cite{NNLO}: in this case, it is found that 
the $O(\assq)$ corrections (NNLO) are of the same size (or even larger) 
of the NLO ones.

Summing up, we can say that on the one 
hand comparison with scaling rules of NRQCD
shows that RG estimates have to be thought of as a lower limit,
while  on the other hand consistency between theory and experiment in total decay rates strongly disfavour much larger colour-octet MEs.
We then conclude that the RG estimates of the colour-octet MEs, 
although sizebly smaller then expected from NRQCD scaling rules, 
are the most reasonable at the present stage of our knowledge.

In fig.~\ref{OCT}  we show in detail the contribution
of the single colour-octet components.
The figure reports LO, direct, and full NLO  contributions for  
states normalized to their respective Born decay widths 
at ${\cal O}(\as \aem)$. Let us consider the $C$-even states first 
($\etah, \chizh, \chith$). 
It is evident that they
contribute to the spectrum with a very similar shape:
there is a strong enhancement at low values of $z$ due to
the fragmentation contribution that is present both
at LO and NLO. Then it is clearly
seen that direct photons mainly contribute near
the end-point, a zone of the spectrum where the fixed-order
calculation is not reliable: in fact there
are clear indications of a need
of resummation both in the short-distance perturbative expansion
in $\as$ and in the long-distance $v$ series.
In ref.~\cite{RW} Rothstein and Wise identified an infinite
class of NRQCD operators, which determine the shape of photonic end-point
functions, and introduced the so-called `shape function',
to be extract from data.  
The overall effect of colour-octet states
would be a smearing of the energy distribution near the end-point on the interval $v^2 \approx 0.1$.
In the case of the $\psih$ component, the direct amplitude is not
divergent in $z=1$ and the NLO correction to the LO 
fragmentation picture is  very small. Indeed the NLO contribution
from direct photons is negative in the $\MSB$-renormalization
scheme and is almost balanced by the other NLO fragmentation terms.

Finally figs.~\ref{fig:plot-res1} and \ref{fig:plot-res2} show the 
total contribution to the spectrum, using the RG estimate for the
non-relativistic matrix elements.  
We notice that the overall effect of octet
states is at its minimum in the central region of the spectrum,
exactly where the singlet LO direct contribution dominates.
This indicates that this region of the spectrum is `safe'
from colour-octet effects, and therefore  we think that it should be used 
to make a comparison with  experimental data. Moreover
this indicates that relativistic corrections to the singlet
(which are indeed important) and higher-order 
strong ones should be included    
to have a consistent theoretical picture at NLO.
On the other side, for small values of $z$, colour-octet
components are not negligible. In this area of the phase space,
the fragmentation components from gluons    
contribute at the same order in $\as$ as the ones from quarks,
and there is no signature to distinguish between the two.
\begin{figure}[t]
\begin{center}
{\unitlength1cm
\begin{picture}(12,10)
\epsfig{file=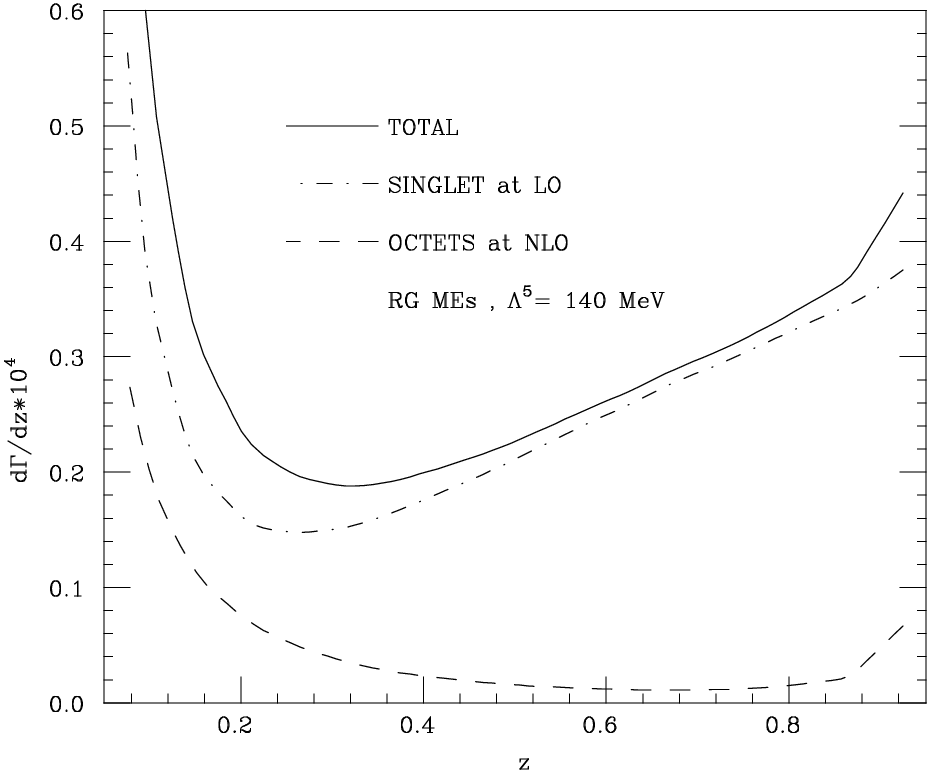,%
          height=10cm,clip=,angle=0}
\end{picture}}
\ccaption{}{\label{fig:plot-res2}
Total colour-octet contribution  on the LO, colour-singlet photon spectrum.
Notice that neither NLO QCD nor relativistic effects are included in the singlet contribution. Normalization and MEs as in fig.~\ref{fig:plot-res1}.}
\end{center}
\end{figure}
Contrary to LO expectations  
in the framework of CSM~\cite{Catani}, we conclude that the decay
of $\Upsilon$ into a photon would not be  useful 
for an estimate of the photon fragmentation functions.

As a final remark, we notice that, not surprisingly, many
of the aspects of the photon spectrum in quarkonium decay, 
resemble those in photoproduction~\cite{Beneke97a,Beneke97b}. Cross-sections plotted
versus the inelasticity $z$ of the quarkonium state show a
very  similar pattern: for $z\approx 1$, a divergence, which is
not supported by the available experimental data, 
reveals the breaking of the NRQCD expansion
in powers of $\as$ and $v$. On the other side, for
low values of $z$, the resolved contributions, which corresponds
to fragmentation in the decays,
are indeed dominated by colour-octet states.

\section{Conclusions}
\label{Conclusions}  

We presented  the calculation of ${\cal O}(\assq \aem)$
colour-octet corrections to the decay of $\Upsilon$ into one 
photon plus light hadrons. Both direct and fragmentation contributions
have been included at NLO. 
In order to study the impact of these contributions on the photon spectrum,
an estimate of the non-perturbative MEs was also given. 
By comparing the available experimental data on fully inclusive and leptonic decay rates with the NLO theoretical predictions of NRQCD, we
found an unexpected result: estimates based on 
na\"\i ve scaling rules result in large colour-octet contributions
to the total rates which are not consistent with the data.
In particular, it turns out that non-perturbative MEs should 
be much smaller then expected from NQRCD scaling rules.      
Nevertheless, using the above mentioned estimates for the non-perturbative
MEs, we showed that there are sizeable effects at the end-points of the spectrum of the photon.
In the case of low values of $z$, the possibility
of measuring the fragmentation function of a gluon into a photon,
which was suggested by the LO  result in the CSM~\cite{Catani}, 
becomes unfeasible: for the colour-octet states both quark and 
gluon fragmentation processes are of the same order in $\as \aem$ 
and there is no signature to distinguish between the two.
Moreover, for values of $z$ near the end-point, breaking of the fixed-order 
calculation is manifest, and the resummations 
of both short-distance coefficient in $\as$ and non-perturbative
MEs in $v$, are called for.    
Nevertheless a `safe' region, for $0.3<z<0.9$, has been found where
octet effects are at their minimum and the perturbative 
expansion in powers of $\as$ and $v$ under proper control.
Following this point of view, we consider the NLO QCD correction  
the colour-singlet differential decay $\rd \Gamma/\rd E_\gamma (\Upsilon\to\threeSone^{[1]}\to\gamma gg)$ worth while to be undertaken. 
\vskip1cm
 
{\bf Acknowledgements.} 
It is a pleasure to thank M.L.~Mangano for valuable advice, discussions
and suggestions during all the stages of this work. We thank
L.~Bourhis for providing us with a ready-to-use set of photon
fragmentation functions. Moreover, we are grateful to M.~Beneke 
and G.T.~Bodwin for reading the manuscript and for their useful suggestions.

\appendix
\section{Symbols and notations}
\label{app:A}
This appendix collects the meaning of various symbols, which are used throughout
the paper.
\\[0.2cm]
\underline{Kinematical factors}:
\be                            
M = 2m \; , \quad\quad
v={\sqrt{1-\frac{M^2}{s}}}\; , \quad
\ee
where $s$ is the partonic centre-of-mass energy squared and $S_{had}$ is the  
hadronic one; $v$ is the velocity of the bound (anti)quark in 
the quarkonium rest frame, $2v$ then being the relative velocity of the quark
and the antiquark.
The following expression is used:
\be                                          
\feps{Q^2} = \left(\frac{4 \pi\mu^2}{Q^2}\right)^{\epsilon}\Gamma(1+\ep) 
= 1+\ep\left(-\gamma_E +\log(4\pi) + \log{\mu^2 \over Q^2}\right) +
  {\cal O}(\epsilon^2) \,.
\ee   
\\[0.2cm]
\underline{Altarelli-Parisi splitting functions}. Several functions related to
the AP splitting kernels enter in our calculations. We collect here our
definitions:
\ba                                             
&& P_{qq}(x) = \cf\left[\frac{1+x^2}{1-x}-\ep(1-x)\right] \,, \\
&&{\cal P}_{qq}(x) =   \cf\left[\frac{1+x^2}{(1-x)_{+}}  + \frac{3}{2}\delta(1-x)\right]\,,\\
&&P_{qg}(x) = \tf\left[x^2+(1-x)^2-2\ep\ x(1-x)\right]\,,\\
&&{\cal P}_{qg}(x) = \tf\left[x^2+(1-x)^2\right]\,,\\
&&P_{\gamma q}(x) = \frac{1+(1-x)^2}{x}- \ep\ x\,,\\
&&{\cal P}_{\gamma q}(x) = \frac{1+(1-x)^2}{x} \,,  \\
&&P_{gq}(x)=\cf\left[\frac{1+(1-x)^2}{x}-\ep\ x\right]\,,\\
&&{\cal P}_{gq}(x)=\cf\left[\frac{1+(1-x)^2}{x}\right]\,,\\
&&P_{gg}(x) = 2\ca\left[\frac{x}{1-x}+\frac{1-x}{x}+x(1-x)\right]\,,\\
&&{\cal P}_{gg}(x) =  2\ca\left[\frac{x}{(1-x)_{+}}+\frac{1-x}{x}+x(1-x)\right]     + \b0\delta(1-x)\,.
\ea
The $P_{ij}$ are the $D$-dimensional splitting
functions that appear in the factorization of collinear singularities from
real emission, while the functions ${\cal P}_{ij}$ are the four-dimensional AP
kernels, which enter in the $\MSB$  collinear counter-terms.
The `+' and `$a$' distributions are defined by:                  
\ba  
   \int_{0}^{1} \, \rd x \; \left[T(x)\right]_+ \phi(x) &=&
   \int_{0}^{1} \, \rd x \; T(x) \; \left[\phi(x)-\phi(1)\right]\,, \\
   \int_{a}^{1} \, \rd x \; \left[T(x)\right]_a \phi(x)&=&
   \int_{a}^{1} \, \rd x \; T(x) \; \left[\phi(x)-\phi(1)\right] \; ,
\ea   
where  $T(x)$ is the function associated to the distributions $[T(x)]_{+,a}$. 
We recall a useful weak distributional identity:
\ba
\left[T(x)\right]_+ = \left[T(x)\right]_a - \delta(1-x)\int_0^a T(x) \rd x\,.
\ea

In particular it is straightforward to get:
\ba
\left(\frac{1}{1-x}\right)_+ &=& \left(\frac{1}{1-x}\right)_a +\delta(1-x)\;\log(1-a)\,,\\
\left(\frac{\log(1-x)}{1-x}\right)_+ &=& \left(\frac{\log(1-x)}{1-x}\right)_a +\delta(1-x)\;\frac{1}{2}\log^2(1-a)\,.
\ea        
\\[0.2cm]
\underline{Colour coefficients}
\ba
\cf = \frac{N_c^2-1}{2\,N_c}\;,\;\;\;\;  \ca = N_c\;,\;\;\;\;  \Bf = \frac{N_c^2-4}{4\,N_c}\;,\;\;\;\; \tf=\frac{1}{2}\,.
\ea
The following standard symbol is used:
\be
\b0 = \frac{11}{6}\ca- \frac{2}{3} \tf\nf\,,
\ee
with $\nf$ the number of flavours {\sl{lighter}} than the bound one.  
\\[0.2cm]
\underline{NRQCD operators}.
To denote a perturbative $\QQ$ state with generic spin and angular momentum
quantum numbers, and in a colour-singlet or colour-octet state,
we use the symbol:                                         
\be
{\cal Q}^{[1,8]}  \equiv \QQ[\spectr^{[1,8]}] \; .
\ee                                               
                                                      
Notice that, according to the discussion in ref.~\cite{Petrelli97}, 
our conventions differ from the Bodwin, Braaten and Lepage ones ~\cite{bbl} 
(labelled here as BBL) in the case of a colour-singlet.
\ba                                       
&&{\oo }_1 = {1\over{2\nc}}  {\o }_1^{\rm BBL}\,, \\
&&{\oo }_8 = {\cal O}_8^{\rm BBL}\,.
\ea                                                                     

\section{Summary of lowest order results}
\label{app:LO}      
\subsection{Born widths}
The  decay rates read
\ba
\Gamma(\ups\to\q^{[1,8]}\to a b) = \hat \Gamma(\q^{[1,8]}\to 
a b)\langle \ups|{\oo }_{[1,8]}(\spectr)|\ups\rangle\, ,
\ea
the short-distance coefficients $\hat\Gamma$ having been 
calculated according to the rules of ref.~\cite{Petrelli97}.
We shall use the short-hand notation
\be
\Gamma(\q^{[1,8]} \to a b ) \equiv \Gamma(\Upsilon  \to \q^{[1,8]} \to a b ) 
\ee                                                                     
to indicate the decay  of the physical quarkonium state
$H$ through the intermediate $\QQ$ state ${\q}^{[1,8]} = \QQ[\spectr^{[1,8]}]$.   
The $D$-dimensional ($D=4-2\epsilon$) $O(\as\aem)$ level decay rates read:
\ba
&&\gbh(\oneSzero^{[8]}\to g\gamma) ={{32 \as \aem \,e^2_Q\, \mu^{4\ep}\pi^2}\over{m^2}}\Phi_{(2)} (1-\ep)(1-2\ep) 
\langle \ups |{\oo}_8(\oneSzero)|\ups\rangle\,,\\
&&\gbh(\threeSone^{[8]}\to g\gamma) =0\,,\\
&&\gbh(\threePzero^{[8]}\to g\gamma) =
{{288\as \aem\,e^2_Q\, \mu^{4\ep}\pi^2}\over{m^4}}
\Phi_{(2)} {{1-\ep}\over{3-2\ep}} 
\langle \ups |{\oo}_8(\threePzero)|\ups\rangle\,, \\
&&\gbh(\threePone^{[8]}\to g\gamma) =0\,,\\
&&\gbh(\threePtwo^{[8]}\to g\gamma) = 
{{64\as \aem\,e^2_Q\, \mu^{4\ep}\pi^2}\over{m^4}}
\Phi_{(2)}
{{(6-13\ep+4\ep^2)}\over{(3-2\ep)(5-2\ep)}}
\langle \ups |{\oo}_8(\threePtwo)|\ups\rangle\,.
\ea
Lowest $O(\assq)$ contributions:
\ba
&&\gbh(\oneSzero^{[8]}\to gg) = \Bf{{16 \assq\, \mu^{4\ep}\pi^2}\over{m^2}}\Phi_{(2)} (1-\ep)(1-2\ep) 
\langle \ups |{\oo}_8(\oneSzero)|\ups\rangle\,,\\
&&\gbh(\threeSone^{[8]}\to \qq) =8\,\frac{\assq\, \mu^{4\epsilon}\pi^2}{m^2}
 \Phi_{(2)}\frac{1-\ep}{3-2\ep}\langle \ups |{\oo}_8(^3S_1)|\ups\rangle\,,\\
&&\gbh(\threePzero^{[8]}\to gg) = \Bf{{144 \assq\, \mu^{4\ep}\pi^2}\over{m^4}}\Phi_{(2)} \frac{(1-\ep)}{(3-2\ep)} 
\langle \ups |{\oo}_8(\threePzero)|\ups\rangle\,,\\
&&\gbh(\threePone^{[8]}\to gg) = 0\,,\\
&&\gbh(\threePtwo^{[8]}\to gg) = \Bf{{32 \assq\, \mu^{4\ep}\pi^2}\over{m^4}}\Phi_{(2)} \frac{(6-13\ep+4\epsilon^2)}{(3-2\ep)(5-2 \epsilon)} 
\langle \ups |{\oo}_8(\threePtwo)|\ups\rangle\,,
\ea
where $\Phi_{(2)}$ is defined according to  eq.~(\ref{eq:phit}).
\subsection{The LO spectrum coefficients $C^{(0)}[\q]$}
We can now read out the lowest-order coefficients according 
to eqs.~(\ref{cexpo})--(\ref{lostru}). For $\q=\chijh\ , \etah$ we have: 
\ba
C^{(0)}_\gamma[\q](z) &=&\;\;\;\Gamma_{\rm Born}[\q\to g\gamma]\delta(1-z)\,,\\
C^{(0)}_g[\q](x) &=& 2\, \Gamma_{\rm Born}[\q\to gg]\delta(1-x)\,,\\
C^{(0)}_q[\q](x) &=&\;\;\;\, 0\,,
\ea
and for $\psih$ :  
\ba
C^{(0)}_\gamma[\psih](z) &=& \;\;\;0\,,\\
C^{(0)}_g[\psih](x) &=& \;\;\;0\,,\\
C^{(0)}_q[\psih](x) &=& \;\;\;\Gamma_{\rm Born}[\q\to \qq]\delta(1-x)\,.
\ea
\section{Summary of  $O(\assq\aem)$ results}
\label{app:NLO}      

\subsection{The NLO photonic coefficients $C^{(1)}_\gamma[\q]$}
We summarize the NLO spectrum coefficient following the convention of  eqs.~(\ref{cexpo})--(\ref{nlostru}).The photon energy fraction is  $z = E_\gamma/m$.
Components $\sim\delta(z)$ have been neglected.
For $\q=\etah ,\chizh , \chith$, we have:
\ba
C^{(1)}_\gamma[\q] &=&\frac{\as}{2\pi} \gb[\q\to g\gamma]
\left[\left(A[{\q}]\;+\; 
2\,\b0\log\frac{\mur}{2m}\right)\delta(1-z)\right.
\nn\\&+& \left.\left(\frac{1}{1-z}\right)_+ f^\gamma_1[{\q}](z)  
+\left(\frac{ \log(1-z)}{1-z}\right)_+  f^\gamma_2[{\q}](z)\right]\,,
\label{photcp}
\ea
where
\ba
&&A[{\etah}] = \cf\left(-10 +\frac{\pi^2}{2} \right) 
  +\ca\left(\frac{121}{18}-\frac{\pi^2}{2}\right) -\frac{10}{9}\nf\tf\,,\\
&&A[{\chizh}] = \cf\left(-\frac{14}{3} +\frac{\pi^2}{2} \right) 
  +\ca\left(\frac{85}{18}-\frac{\pi^2}{2}\right)-\frac{10}{9}\nf\tf \,,\\
&&A[{\chith}] = -8\cf +\ca\left(\frac{47}{9}+\log 2 \right)
-\frac{8}{45}\nf\tf\,,
\ea
and 
\ba
&&f^\gamma_1[{\etah}](z)  =  \ca{{\,\left( -72 + 144\,z - 176\,{z^2} + 104\,{z^3} 
  - 23\,{z^4} \right)}\over {6\,{{\left( -2 + z \right) }^2}\,z}} 
  + \nf\tf\frac{2}{3} z\,,\\
&&f^\gamma_1[{\chizh}](z) =\ca\frac{1}{54\,{{\left( -2 + z \right) }^4}\,{z^3}} 
  \,\left( -960 + 3360\,z - 6224\,{z^2} + 5312\,{z^3} - 1544\,{z^4} - 520\,
  {z^5}\right.\nn\\&&\left.\;\;\;\;\;\;\;\; + 496\,{z^6} - 136\,{z^7} 
  + 9\,{z^8} \right)+\nf\tf\frac{2}{27\,z}{(z+2)^2}\,,\\
&&f^\gamma_1[{\chith}](z) = \ca \frac{1}{36\,{{\left( -2 + z \right) }^4}\,{z^3}} 
  \,\left( -240 + 1848\,z - 7820\,{z^2} + 13976\,{z^3} - 12710\,{z^4} 
  + 6254\,{z^5} \right.\nn\\&&\left.\;\;\;\;\;\;\;\;- 1628\,{z^6} 
  + 197\,{z^7} - 15\,{z^8}
  \right)  + \nf\tf \frac{1}{9\,z}\left(10-5z +z^2\right)\,,\\
&&f^\gamma_2[{\etah}](z)  =  \ca{{2\,\,\left(+ 12 - 36\,z + 56\,{z^2} - 52\,{z^3} 
  + 28\,{z^4} -  8\,{z^5} + {z^6} \right) }\over {{{\left( -2 + z \right) }^3}
  \,{z^2}}} \,, \\
&&f^\gamma_2[{\chizh}](z) =  \ca\frac{2}{9\,\left( -2 + z \right)^5\,z^4 } 
  \,\left(+ 160 - 720\,z + 1624\,{z^2} - 2016\,{z^3} + 1360\,{z^4} 
  - 468\,{z^5}\right.\nn\\&&  \left.\;\;\;\;\;\;\;\;+ 104\,{z^6} 
  - 66\,{z^7} + 40\,{z^8} - 10\,{z^9} + {z^{10}} \right)\,, \\
&&f^\gamma_2[{\chith}](z) =\ca \frac{1}{3\,\left( -2 + z \right) ^5\,{z^4}} 
  \,\left(+ 40 - 348\,z + 1618\,{z^2} - 3684\,{z^3} + 4702\,{z^4} 
  - 3669\,{z^5}\right.\nn\\ &&\left.\;\;\;\;\;\;\;\;+ 1826\,{z^6} 
  - 582\,{z^7} + 115\,{z^8} - 13\,{z^9} + {z^{10}} \right)\,.
\ea
For the $\psih$ component we get:
\ba 
C_\gamma^{(1)}[\psih]  
&=&\frac{20 \aem e_Q^2 \assq}{9}
\left[ \frac{1}{z(-2+z)^2}\left(8-12 z + 7 z^2 - 2 z^3\right)\right.\nn\\
&+&\left.\frac{2}{(-2+z)^3 z^2}(-1+z)(8-12 z+5 z^2)\log(1-z)\right]
\frac{\langle\ups\vert{\cal O}_8[\threeSone]\vert \ups\rangle}{m^2}\nn\\
&+& \Gamma_{\rm Born}(\psih\to\qq)\frac{\aem}{\pi}
P_{\gamma q}(z)\,\left(\log \frac{4m^2}{\mufrags} + \log (1 - z) + 2\,\log z \right)\sum_q e_q^2 \,,\nn\\
\label{photcm}
\ea
and finally for $\chioh$:
\ba 
C_\gamma^{(1)}[\chioh]   
&=& \frac{2 \aem e_Q^2\assq}{3}\left[ \frac{1}{(-2+z)^4z^3}
(240+312 z -2620 z^2 + 4204 z^3 -3150 z^4 + 1260z^5\right.\nn\\
&-& 276 z^6 + 31 z^7)
+\frac{12}{(-2+z)^5z^4}(-1+z)
( 40 + 52 z - 430 z^2 + 716 z^3 - 588 z^4\nn\\     
&+&\left. 275 z^5 - 74 z^6 + 11 z^7 - z^8)\log(1-z)+ 
\frac{2}{3}\nf\frac{2-x}{x} \right]\frac{\langle \ups\vert{\cal O}_8[\threePone]\vert\ups\rangle}{m^4} \,.
\ea

\subsection{The NLO gluonic coefficients $C_g^{(1)}[\q]$}
In this section we present the NLO QCD  spectrum of the gluon arising from the colour-octet components. Contributions $\sim\delta(x)$ have been neglected.
The gluon energy fraction is denoted by $x= E_g/m$ .
For $\q=\etah ,\chizh , \chith$,  we have:
\ba 
C_g^{(1)}[\q]   
&=&\frac{\as}{\pi}\gb[\q\to gg]
\left[ \log \frac{4m^2}{\mufrags} {\cal P}_{gg}(x)+ 2 \log x 
{\cal P}_{gg}(x) + \left(\frac{\log(1-x)}{1-x}\right)_+ (1-x)P_{gg}(x)\right.\nn\\
 &+&\left.  \left(\frac{1}{1-x}\right)_+ f[\q](x) + \left( B[\q] + 4 \b0\log\frac{\mur}{2m}\right) \delta(1-x)\right]\,,\nn\\ 
\ea
where:
\ba
B[\etah] &=& \cf\left(-10+\frac{\pi^2}{2}\right) +\ca\left(\frac{139}{18}-\frac{1}{12}\pi^2\right) -\frac{10}{9}\nf\tf\\
B[\chizh]&=& \cf\left(-\frac{14}{3} +\frac{\pi^2}{2}\right) + \ca\left(\frac{235}{54}+\frac{70}{27}\log 2 -\frac{1}{12}\pi^2\right) -\frac{10}{9}\nf\tf\\
B[\chith]&=& -8\cf + \ca\left( 5 +\frac{14}{9}\log 2 -\frac{1}{6}\pi^2\right) -\frac{8}{45}\nf\tf\,,
\ea
and furthermore
\ba
f_g[\etah](x) &=&  \frac{\ca}{6\,{{\left( -2 + x \right) }^2}\,x}( -120 + 336\,x - 494\,{x^2} + 410\,{x^3} 
- 215\,{x^4} + 72\,{x^5} - 12\,{x^6})\nn\\
&+&\frac{2\,{\ca}\,\left( -1 + x \right)}{{{\left( 2 - x \right) }^3}\,{x^2}}
      ( 16 - 40\,x + 50\,{x^2} - 26\,{x^3} - 8\,{x^4} + 16\,{x^5} -
        7\,{x^6} + {x^7}) \,\log (1 - x)\nn\\
&+&\nf \tf\frac{2}{3}x\,,\\
&&\nn\\
f_g[\chizh](x) &=&\frac{\ca}{54\,{{\left( -2 + x \right) }^4}\,{x^3}}\,( -1536 + 5376\,x - 9632\,{x^2} + 10016\,{x^3} - 9288\,{x^4} + 12976\,{x^5} \nn\\ &-&  16906\,{x^6} +13918\,{x^7} - 6623\,{x^8} + 1664\,{x^9} - 172\,{x^{10}})\nn\\
&+& \frac{2\,{\ca}\,\left( -1 + x \right)}{9\,{{\left( 2 - x \right) }^5}\,{x^4}} \,( 256 - 896\,x + 1504\,{x^2} - 1008\,{x^3} - 516\,{x^4} +
        1792\,{x^5} - 2276\,{x^6}\nn\\ 
&+& 2011\,{x^7} - 1220\,{x^8} + 464\,{x^9} -
        99\,{x^{10}} + 9\,{x^{11}}) \,\log (1 - x) \nn\\
&+&\nf \tf\frac{2(2+x)^2}{27x}\,,\\
&&\nn\\
f_g[\chith](x)&=& \frac{\ca}{36\,{{\left( -2 + x \right) }^4}\,{x^3}} \,( -384 + 3072\,x - 12704\,{x^2} + 25376\,{x^3} -
        30738\,{x^4} + 26998\,{x^5}\nn\\ 
&-&19231\,{x^6} +10924\,{x^7} -  4373\,{x^8} + 1028\,{x^9}-106\,{x^{10}})\nn\\
&+&\frac{ {\ca}\,\left( -1 + x \right)}{3\,{{\left( 2 - x \right) }^5}\,{x^4}} 
 ( 64 - 512\,x + 2032\,{x^2} - 3840\,{x^3} + 3747\,{x^4} -
        1577\,{x^5} - 515\,{x^6}\nn\\ 
&+& 1162\,{x^7} - 800\,{x^8} + 308\,{x^9} -
        66\,{x^{10}} + 6\,{x^{11}}) \,\log (1 - x)\nn\\
&+&\nf\tf\frac{1}{9x}\left(10 -5x+x^2\right)\,.
\ea
For the $\psih$ component we get:
\ba 
&&C_g^{(1)}[\psih] = 
  \frac{\ascube}{18}\left[
  \frac{1}{(-2+x)^2 x}(1168 - 3264 x + 3740 x^2 - 2200 x^3 + 693 x^4 - 108 x^5)
  \right.\nn\\&+& 
  \left.\frac{4(-1+x) }{(-2+x)^3 x^2}(584 -1632 x + 1904 x^4 -
  1134 x^3 + 324 x^4 -27 x^5)\log (1-x)\right]
  \frac{\langle \ups\vert\oo_8[\threeSone]\vert \ups\rangle}{m^2}\nn\\
&+& \nf \Gamma_{\rm Born}(\psih\to\qq) \frac{\as}{\pi}\left[
  {{-3 + 3\,x - \,{x^2} }\over {x}} + 
  \left(\log \frac{4m^2}{\mufrags} + \log (1 - x) +
  2\,\log x \right)P_{gq}(x)\right]\,.\nn\\
\ea
Finally for $\chioh$:
\ba 
&&C_g^{(1)}[\chioh] =
\frac{5\ascube}{18}\left[ \frac{1}{(-2+x)^4x^3}
(384 + 384 x - 4192 x^2 + 7552 x^3 - 6446 x^4 + 2876 x^5 \right.\nn\\
&-& 485 x^6 -141 x^7 + 82 x^8 - 10 x^9 )
+\frac{12}{(-2+x)^5x^4}(-1+x)(64 + 64x - 688x^2 + 1280 x^3\nn\\
&-&\left. 1181 x^4 + 626 x^5 - 195 x^6 + 36 x^7 - 4 x^8 )\log(1-x) 
+ \frac{2}{3} \nf\frac{2-x}{x}\right]\frac{\langle \ups\vert\oo_8[\threePone]\vert \ups\rangle}{m^4}\,.
\ea
\subsection{The NLO quark coefficients $C_q^{(1)}[\q]$ }
We report in this section  the quark energy spectrum in $\q\to\qq\,g$ decays.
The adimensional energy of the quark $E_q/m$ is denoted by $x$.
\ba
C_q^{(1)}[\etah]  
&=&\frac{\as}{\pi}\gb[\etah\to gg]
\left[{\cal P}_{qg}(x)\log \frac{4m^2}{\mufrags}\right.\nn\\
&+& \left. 2\,x(1-x)\tf +{\cal P}_{qg}(x) \log[x^2(1-x)] + f[\etah](x)\right]\,,
\ea
where 
\ba
f_q[\etah](x) = x(1-x)(1+\log(1-x))\,.
\ea
We have
\ba
C_q^{(1)}[\psih] 
&=&\gb[\psih\to\qq]\frac{\as}{\pi}
\left[\frac{1}{2} {\cal P}_{q q}(x)\log \frac{4m^2}{\mufrags}
\right.  \nn\\ 
&+& \cf \frac{1+x^2}{(1-x)_+}\log\,x + \frac{\cf}{2}\,(1-x) + 
\frac{1}{2}(1-x)\left(\frac{\log(1-x)}{1-x}\right)_+ P_{q q}(x)\nn\\ 
&+& \left.\left(\frac{1}{1-x}\right)_+ f_q[\psih](x) + A[\psih]\delta(1-x)\right]\,,
\ea
where
\ba
A[\psih] = \cf\left(-\frac{25}{4}+\frac{\pi^2}{3}\right) +\ca\left( \frac{50}{9} +\frac{2}{3}\log\,2 -\frac{\pi^2}{4}\right) -\frac{10}{9}\nf\tf+2\,\log\frac{\mur}{2m}\,,
\ea
and
\ba
f_q[\psih](x) = \cf\frac{x}{4}\left(-4+x\right) -\ca\frac{x}{2}\left(5-5x+2x^2\right)+\ca(1-x)(-2+x)\log(1-x)\,
\ea

\ba
&&C_q^{(1)}[\chijh]= 
\Bf\ascube\delta(1-x)\left[ -\frac{8}{9}\log\frac{\mul}{2m} + a_J\right]\frac{\langle \ups\vert{\cal O}_8[\threePJ]\vert \ups\rangle}{m^4}\nn\\&& + \frac{\as}{\pi}\Gamma_{\rm Born}[\chijh\to g g]\left[ 2 x (1-x)\tf + \log [x^2
(1-x)] {\cal P}_{qg}(x) +\left(\frac{1}{1-x}\right)_+ f_q^{(J)}(x)\right.\nn\\
&& 
\left.+ \,{\cal P}_{qg}(x)\log \frac{4m^2}{\mufrags} \right]\,, \quad[J=0,2] \,.
\ea
We also have 
\ba
&&C_q^{(1)}[\chioh]= 
 \Bf\ascube\delta(1-x)\left[ -\frac{8}{9}\log\frac{\mul}{2m} + a_1\right]\frac{\langle \ups\vert{\cal O}_8[\threePone]\vert \ups\rangle}{m^4}\nn\\ 
&&+\ascube\Bf \left(\frac{1}{1-x}\right)_+ f_q^{(1)}(x) \frac{\langle \ups\vert{\cal O}_8[\threePone]\vert \ups\rangle}{m^4}\,,
\ea
where
\ba
a_0 = \frac{2}{9}\,,\;\;\; a_1 = \frac{1}{9}\,,\;\;\;\; a_2 = \frac{7}{45}\,,
\ea
and finally
\ba
f_q^{(0)}(x) &=&\frac{1}{27}\left[ x(33 - 72\,{x} + 43\,{x^2}) - 3(1-x)(4-9 x + 9 x^2)\log (1 - x)\right]\,,\\
f_q^{(1)}(x) &=& \frac{2}{9}\left[ x(3+6 x -5 x^2) +3(1-x)\log(1-x)\right]\,, \\
f_q^{(2)}(x)&=& \frac{1}{36}\left[ x(57-90 x + 53 x^2)- 3 (1-x)(5-12 x + 12 x^2)\log(1-x)\right]\,.
\ea


\vfill
\end{document}